\documentclass[useAMS,usenatbib]{mn2e}
\usepackage{amsmath}
\usepackage{mathabx}
\usepackage{dsfont}
\usepackage{graphicx}
\usepackage{mathtools}

\title[Accretion, radial flows and abundance gradients]{Accretion, radial flows and abundance gradients in spiral galaxies}
\author[G. Pezzulli and F. Fraternali]{Gabriele Pezzulli$^{1}$\thanks{E-mail:
gabriele.pezzulli@unibo.it} and Filippo Fraternali$^{1, 2}$ \\
$^{1}$ Department of Physics and Astronomy, University of Bologna, Viale Berti Pichat 6/2, I-40127 Bologna, Italy \\
$^{2}$ Kapteyn Astronomical Institute, University of Groningen, Postbus 800, NL-9700 AV Groningen, The Netherlands \\}

\begin{document}

\date{Accepted 2015 Oct 13.
Received 2015 Oct 13;
in original form 2015 July 23
}

\pagerange{\pageref{firstpage}--\pageref{lastpage}} \pubyear{2015}

\maketitle

\label{firstpage}

\begin{abstract}
The metal-poor gas continuously accreting on to the discs of spiral galaxies is unlikely to arrive from the intergalactic medium (IGM) with exactly the same rotation velocity as the galaxy itself and even a small angular momentum mismatch inevitably drives radial gas flows within the disc, with significant consequences to galaxy evolution. 
Here we provide some general analytic tools to compute accretion profiles, radial gas flows and abundance gradients in spiral galaxies as a function of the angular momentum of the accreting material. 
We generalize existing solutions for the decomposition of the gas flows, required to reproduce the structural properties of galaxy discs, into direct accretion from the IGM and a radial mass flux within the disc. 
We then solve the equation of metallicity evolution in the presence of radial gas flows with a novel method, based on characteristic lines, which greatly reduces the numerical demand on the computation and sheds light on the crucial role of boundary conditions on the abundance profiles predicted by theoretical models. We also discuss how structural and chemical constraints can be combined to disentangle the contributions of inside-out growth and radial flows in the development of abundance gradients in spiral galaxies. 
Illustrative examples are provided throughout with parameters plausible for the Milky Way. 
We find that the material accreting on the Milky Way should rotate at $70-80$ per cent of the rotational velocity of the disc, in agreement with previous estimates.
\end{abstract}


\section{Introduction}\label{sec::Introduction}
Continuous accretion of metal-poor gas is known to be a crucial ingredient in the evolution of spiral galaxies, from a wide variety of evidence (e.g.\ \citealt{PP75}; \citealt{FT12}; \citealt{Rhee+13}; \citealt{Sanchez-AlmeidaReview} and references therein). Unfortunately, direct observations of the actual accretion process has proven to be very elusive (e.g.\ \citealt{Sancisi+08}; \citealt{DiTeodoro} and references therein), with the consequence that the spatial distribution of the accretion is largely unknown, even for our own Galaxy at the present time, let alone for external galaxies and on cosmological time-scales. Hence, indirect inference from observed properties of galaxies is still necessary.

Models for the evolution of spiral galaxies, and of our own Galaxy in particular, have always taken great advantage from simple parametrizations of the accretion rate surface density as a function of time and radius (e.g.\ \citealt{Pagel2009}; \citealt{Matteucci2012} and references therein). In order to constrain this function from observations, a powerful and widely used approach is the requirement, for every annulus with given galactocentric radius $R$, that the integral over time of the accretion rate surface density at $R$ equals the total present-time observed baryonic surface density there.

One problem of the above method is that it implicitly assumes that the material accreting on a galaxy disc at some radius will keep staying there for the whole galactic history (the so-called \emph{independent annuli} approach). This occurrence is in general very unlikely, being in contradiction with a basic dynamical requirement, namely angular momentum conservation. In fact, as it was first emphasized by \cite{MV81}, whenever the specific angular momentum of the accreting material is not exactly equal to that of the disc in the point of impact (which would be a fine-tuned coincidence), radial flows will inevitably onset within the disc in order for total angular momentum to be conserved. This implies the break of the independent annuli assumption and, in particular, of the link between the total mass in a given annulus and the integral over time of the mass accretion rate there.

A direct detection of net radial motions in the gaseous discs of spirals would be very valuable, since it would give constraints on the dynamics of the accretion process on to spiral galaxies. 
Unfortunately, the velocities predicted as a consequence of the aforementioned effect are so low that they are barely measurable and, most importantly, it is still not clear how they could be observationally distinguished from spurious effects driven by non-axisymmetric disturbances (\citealt{WongBlitzBosma04}). Therefore, indirect methods are necessary, which are able to capture the integrated signal of such low velocities across the whole history of a galaxy disc. Chemical evolution provides the desired alternative, since, at any given time, the distribution of metals in the interstellar medium (ISM) depends on the whole history of enrichment and dilution that each gas element has undergone throughout its path across the galaxy.

A considerable amount of work has been devoted to the effect of radial flows on chemical evolution, and in particular on the development of abundance gradients (see e.g.\ \citealt{LF85}; \citealt{GK92}; \citealt{PC00}; \citealt{Cavichia+14}; \citealt{KPA15}, among many others), sometimes in conjunction with other mechanisms, like inside-out formation and radial variations (either smooth or discontinuous) in the normalization of the star formation law (e.g.\ \citealt{Clarke89}; \citealt{ChamchamTayler94}; \citealt{SpitoniMatteucci2011}), both in the Local Universe and at high redshift (\citealt{Mott+13}). Several mechanisms can in principle contribute to the onset of ordered radial gas flows within the discs of spirals (e.g.\ \citealt{LF85}; \citealt{ThonMeusinger98}) and turbulence can cause additional transport of metals (e.g.\ \citealt{YK12}; \citealt{Petit+15}), though order of magnitude estimates suggest angular momentum deficit in accretion to be the dominant process, or one of the dominant processes, in this respect (\citealt{BS12}). However, only few works, to our knowledge, have so far implemented radial flows that are consistent with angular momentum conservation.

In their seminal work, \cite{MV81} considered the extreme case of infall of gas completely devoid of angular momentum. \cite{LF85} improved the angular momentum equation allowing for non-vanishing rotation in the accreting material, but they did not use it in their models, preferring simpler parametrizations of the radial velocity.

Two major improvements came in the field with the work by \cite{PT89}, though both of them maybe received somewhat less consideration than they were worth of. First, these authors introduced the fundamental concept of \emph{effective accretion}, which is the amount of gas that has to come at a given annulus at a given time in order to sustain a given structural evolution. The effective accretion, rather than the accretion itself, is the quantity that can be directly constrained by the present-day structure of galaxies (e.g. its integral over time equals the present-day baryonic surface density) and which should, therefore, be preferentially chosen as a starting point for semi-analytic models of galaxy evolution. In general, the effective accretion needs to be decomposed in two contributions, direct accretion from the intergalactic medium (IGM) and an internal contribution due to radial flows, a problem which we refer to as \emph{mass flux decomposition}. To better appreciate the importance of this, it may be useful to consider that fixing an accretion profile, rather than the profile of effective accretion, in a set of models with radial flows, has the consequence that different models will have a different final structure (e.g. models with inward radial flows will be more centrally concentrated) and therefore they cannot all succeed in satisfying observational structural requirements (e.g.\ \cite{LF85}; \cite{Tosi88}; \cite{GK92}). The approach based on effective accretion, instead, guarantees that all the models match the same structural constraints and therefore allows to validate or discard radial flow models on the basis of chemical information. Note that in this way the accretion profile is a prediction, rather than an input, of a model with radial flows.

As a second improvement, \cite{PT89} provided some analytic solutions to the mass flux decomposition problem as a function of the angular momentum of the infalling material. Their solutions, however, were valid only for some very special values of the angular momentum and did not allow a continuum of models to be explored.

The mass flux decomposition problem has been recently reconsidered, from a numerical point of view, by \cite{BS12}, who solved a discretized version of the equation of angular momentum conservation and were able to compute a fine grid of models. Implementing their technique into the detailed Milky Way evolution model by \cite{SB09}, they provided the current benchmark for the dynamical properties of accreting material on our own Galaxy.

The aim of the present work is to develop the analytic approach to the problem. To this purpose, we provide some new analytic methods to compute gas flows and abundance gradients in models of spiral galaxy evolution, under general conditions. The paper is organized as follows. In Sec.\ \ref{sec::MinimalModel}, a minimal model for the evolution of spiral galaxies is discussed and some consequences are investigated of the independent-annuli assumption on the chemical evolution of spiral galaxies. In Sec.\ \ref{sec::AngularMomentumAccretion}, the general analytic solution is given to the mass flux decomposition problem, as a function of the angular momentum of the accreting material. In Sec.\ \ref{sec::ChemicalEvolutionRadialFlows}, a novel method, based on characteristic lines, is proposed to compute chemical evolution in the presence of radial flows and its importance is elucidated to trace the effect of boundary conditions on the abundance profiles in the outskirts of galaxy discs. In Sec.\ \ref{sec::InsideOutModel}, we show how structural and chemical information can be combined to disentangle the effects of inside-out growth and radial flows on the development of abundance gradients in spiral galaxies. Though the main rationale is generality, examples are provided throughout, calibrated on Milky Way like parameters, to allow for a comparison with observations and with previous work on the subject.

\section{A minimal model for the evolution of galaxy discs}\label{sec::MinimalModel}
Before looking at more complex situations, we consider in this Section the simplest possible model for the evolution of a galaxy disc. Without any presumption of detail, we just make elementary assumptions, based on few general properties of the discs of spiral galaxies and simple enough to allow for an analytic description. We use dimensionless units when possible, but also give an illustrative example with parameters chosen to be plausible for the Milky Way.

\subsection{Exponential discs obeying the Kennicutt-Schmidt law}\label{sec::ExpKennicuttDisc}
As basic structural requirements, we ask that stellar discs have an exponential radial mass distribution:
\begin{equation}\label{expdisc}
\Sigma_\star(t, R) = \frac{M_\star(t)}{2 \pi R_\star^2} e^{-\frac{R}{R_\star}}
\end{equation}
with obvious meaning of symbols, and form stars according to the Kennicutt-Schmidt law:
\begin{equation}\label{Klaw}
\dot{\Sigma}_\star = A \Sigma_\textrm{g}^N
\end{equation}

The fiducial parameters of the Kennicutt-Schmidt law are $N = 1.4$ and $A = 0.11375$, if surface densities are measured in $\textrm{M}_\Sun \; \textrm{pc}^{-2}$ and times in $\textrm{Gyr}$. These values are taken from \cite{K98}, with $A$ corrected for a helium factor equal to 1.36 and then multiplied by a factor $(1-\mathcal{R})$, with an assumed \emph{return fraction} $\mathcal{R} = 0.3$, to take into account material returned from stars to the ISM during stellar evolution. Note that this implies that our $\dot{\Sigma}_\star$ represents the net (or \emph{reduced}) star formation rate surface density (SFRD), rather than the instantaneous one.

Equations \eqref{expdisc} and \eqref{Klaw} imply that also the gas distribution is exponential:
\begin{equation}
\Sigma_\textrm{g}(t, R) = \frac{M_\textrm{g}(t)}{2 \pi R_\textrm{g}^2} e^{- \frac{R}{R_\textrm{g}}}
\end{equation}
with a scalelength that is larger than the one of the stars, according to $R_\textrm{g} = NR_\star$, while the total gas mass $M_\textrm{g}$ is linked to the global star formation rate $\dot{M}_\star$ by the equation:
\begin{equation}\label{globalKlaw}
\dot{M}_\star = \hat{A} M_\textrm{g}^N
\end{equation}
with:
\begin{equation}
\hat{A} \vcentcolon= \frac{A}{N^2(2 \pi R_\textrm{g}^2)^{N-1}}
\end{equation}
Equation \eqref{globalKlaw} resembles in form the Kennicutt-Schmidt law, with the important differences that it refers to masses, rather than surface densities, and its normalization $\hat{A}$ is not universal, but a function of the scalelength. For a fiducial stellar scalelength $R_\star = 2.5 \; \textrm{kpc}$, which should be similar to the value appropriate for our Galaxy, it is $R_\textrm{g} = 3.5 \; \textrm{kpc}$ (see also \citealt{SB09}) and therefore $\hat{A} = 0.16187$, if masses are measured in units of $10^9 \; \textrm{M}_\Sun$ and times in Gyr.

Finally, we consider the simple case of an exponentially declining star formation history (SFH):
\begin{equation}\label{expSFH}
\dot{M_\star}(t) = \frac{M_{\star, \infty}}{t_\star} e^{- \frac{t}{t_\star}}
\end{equation}
where $M_{\star, \infty}$ is the asymptotic value of the stellar mass and $t_\star$ is the star formation decline time-scale. Equations \eqref{expSFH} and \eqref{globalKlaw} imply that the gaseous mass is also exponentially declining with time, but with a larger time-scale $t_\textrm{g} = N t_\star$ and an initial value $M_{\textrm{g}, 0}$ linked to the other parameters by the relation:
\begin{equation}\label{Mstarinfrel}
M_{\star, \infty} = \hat{A} M_{\textrm{g}, 0}^{N} t_\star
\end{equation}

The two independent parameters of the model, $M_{\star, \infty}$ and $t_\star$, can be uniquely constrained, for a given galaxy, from the global properties of that galaxy at the present time. In particular, the quantity $x = t_0/t_\star$, $t_0$ being the age of the disc, can be inferred from observations by inverting the relation:
\begin{equation}
\frac{x}{e^x - 1} = \frac{t_0 \hat{A} M_\textrm{g}^N(t_0)}{M_\star(t_0)}
\end{equation}
and $M_{\star, \infty}$ can then be found by means of equation \eqref{Mstarinfrel}. In our fiducial example, we consider a $12 \; \textrm{Gyr}$ aged disc with a stellar mass $M_\star = 4 \times 10^{10} \textrm{M}_\Sun$ and a gaseous mass $M_\textrm{g} = 6 \times 10^9 \; \textrm{M}_\Sun$. The resulting parameters of the model are $t_\star = 12.5 \; \textrm{Gyr}$ and  $M_{\star, \infty} = 6.5 \times 10^{10} \; \textrm{M}_\Sun$.

\subsection{Effective accretion rate surface density}\label{sec::EffectiveAccretion}
Following \cite{PT89}, we define the \emph{effective} accretion rate surface density as:
\begin{equation}\label{dotSigmaEffdef}
\dot{\Sigma}_\textrm{eff} (t, R) \vcentcolon= \frac{\partial}{\partial t} (\Sigma_\star + \Sigma_\textrm{g}) (t, R)
\end{equation}
which is the amount of gas that is needed at a given time $t$ and a given radius $R$ in order to sustain a given structural evolution. This material can either directly come from the IGM, or through the disc via radial flows, or, in general, from a combination of the two routes (\citealt{PT89}; \citealt{SB09}). This fact is conveniently formalized by the equation of conservation of mass. If, in the evolution of the gas component, a sink term $\dot{\Sigma}_\star$ and a source term $\dot{\Sigma}_\textrm{acc}$ are accounted for, to describe star formation and accretion, respectively, then the continuity equation can be written:
\begin{equation}\label{continuity}
\dot{\Sigma}_\textrm{eff} = \dot{\Sigma}_\textrm{acc} - \frac{1}{2 \pi R} \frac{\partial \mu}{\partial R}
\end{equation}
where:
\begin{equation}\label{defmu}
\mu \vcentcolon= 2 \pi R \Sigma_\textrm{g} u_R
\end{equation}
is the radial gaseous mass flux, $u_R$ being the net radial velocity of the gas.

If an independent annuli-approach is adopted, then radial flows are neglected, the second term in the r.h.s. of equation \eqref{continuity} vanishes and therefore $\dot{\Sigma}_\textrm{acc} = \dot{\Sigma}_\textrm{eff}$. In general, however, a different accretion profile is needed, a problem which we will come back to in Sec. \ref{sec::MassFluxDec}.

All the considerations above are fully general and do not dependent on the particular structural evolution model under consideration. In the particular case of an exponential disc obeying the Kennicutt-Schmidt law (Sec.\ \ref{sec::ExpKennicuttDisc}), equation \eqref{dotSigmaEffdef} reads:
\begin{equation}\label{dotSigmaEffExp}
\dot{\Sigma}_\textrm{eff} = \frac{\dot{M}_\star(t)}{2 \pi R_\star^2} e^{-\frac{R}{R_\star}} + \frac{\dot{M}_\textrm{g}(t)}{2 \pi R_\textrm{g}^2} e^{-\frac{R}{R_\textrm{g}}}
\end{equation}
where all the functions and parameters are as specified in Sec. \ref{sec::ExpKennicuttDisc}. As long as radial flows are neglected, equation \eqref{dotSigmaEffExp} gives the accretion profile at any time in our minimal model.

\subsection{Metallicity evolution with independent annuli}\label{sec::IndependentAnnuliMetalEvol}
In a model with independent annuli, the metallicity evolution of each annulus is governed by the equation:
\begin{equation}\label{metaleq0}
\frac{\partial  \tilde{X}_i}{\partial t} = \frac{\dot{\Sigma}_\star}{\Sigma_\textrm{g}} - \tilde{X}_i \frac{\dot{\Sigma}_\textrm{acc}}{\Sigma_\textrm{g}}
\end{equation}
where $\tilde{X}_i = X_i/y_i$ is the abundance by mass of an element $i$, normalized on its yield $y_i$. Strictly speaking, equation \eqref{metaleq0} is an approximation, because, consistently with out treatment of star formation (see Sec. \ref{sec::ExpKennicuttDisc}), it assumes instantaneous recycling. For this reason, it can be considered reliable only to predict the abundances of $\alpha$ elements, which are produced by short-lived stars. Furthermore, equation \eqref{metaleq0} can only be straightforwardly applied to abundances in the ISM; the spatial distribution of stellar abundances, in fact, is known to be severely affected by stellar radial migration (e.g.\ \citealt{SB09}; \citealt{KPA13}; \citealt{MCM14}). However, restricting our attention to $\alpha$ elements in the ISM will not prevent us to draw significant conclusions.

If $\tilde{X}_i = 0$ at $t = 0$, the explicit solution for equation \eqref{metaleq0} is:
\begin{equation}\label{metalsol}
\tilde{X}_i(t, R) = e^{-\sigma(t, R)} \int_0^t e^{\sigma(t', R)} \frac{\dot{\Sigma}_\star}{\Sigma_\textrm{g}}(t', R) dt'
\end{equation}
where we introduced the dimensionless coordinate:
\begin{equation}\label{defsigma}
\sigma(t, R) \vcentcolon=  \int_0^t \frac{\dot{\Sigma}_\textrm{acc}}{\Sigma_\textrm{g}}(t', R) dt'
\end{equation}

Notice that $\sigma$ increases with time at different rates with varying $R$, implying different time-scales for chemical evolution at different radii.

\subsection{Independent-annuli chemical evolution of exponential discs}\label{sec::Conundrum}
The independent-annuli metallicity evolution of the minimal model of Sec. \ref{sec::ExpKennicuttDisc} is conveniently computed in terms of the dimensionless time and space coordinates:
\begin{equation}
\tau \vcentcolon= \frac{t}{t_1} \qquad \rho \vcentcolon= \frac{R}{R_1}
\end{equation}
where:
\begin{equation}
t_1 \vcentcolon= \frac{N}{N-1} t_\star \qquad R_1 \vcentcolon= \frac{N}{N-1} R_\star
\end{equation}
In our fiducial model, $t_1 =  43.8 \; \textrm{Gyr}$ and $R_1 = 8.75 \; \textrm{kpc}$.

Under the independent-annuli assumption, $\dot{\Sigma}_\textrm{acc}$ is given by equation \eqref{dotSigmaEffExp} (see Sec. \ref{sec::EffectiveAccretion}), and therefore the coordinate $\sigma$ defined in equation \eqref{defsigma} is:
\begin{equation}\label{sigmaspecialcase}
\sigma(\tau, \rho) = q_0 e^{- \rho}(1 - e^{-\tau}) - \frac{\tau}{N-1}
\end{equation}
where $q_0 = \hat{A} t_1 M_{\textrm{g}, 0}^{N-1}/N^2$, which is a dimensionless parameter. In the fiducial model, $q_0 = 37.5$

Note that the coordinate \eqref{sigmaspecialcase} is time-increasing (and thus, well defined) for $\tau + \rho < \ln((N-1)q_0)$. The reason is that outside this domain equation \eqref{dotSigmaEffExp} formally describes an effective wind and therefore cannot be used as an accretion term into equation \eqref{metaleq0} anymore. Our fiducial model matches the requirement for its present age out to more than $20 \; \textrm{kpc}$, which is large enough for a comparison with observations. Within this domain, the metallicity evolution \eqref{metalsol} reads:
\begin{equation}\label{metalsolexp}
\tilde{X}_i (\tau, \rho) = q_0 e^{-\sigma(\tau, \rho) - \rho} \int_0^\tau e^{\sigma(\tau', \rho)-\tau'} d \tau'
\end{equation}
with $\sigma$ given by equation \eqref{sigmaspecialcase}.

Equation \eqref{metalsolexp} gives the general evolution of the abundance profiles for an exponential disc obeying the Kennicutt-Schmidt law with an exponentially declining SFH, in the absence of radial flows. It provides a family of self-similar solutions for each value of the dimensionless parameter $q_0$.

\begin{figure}
\centering
\includegraphics[width=9cm]{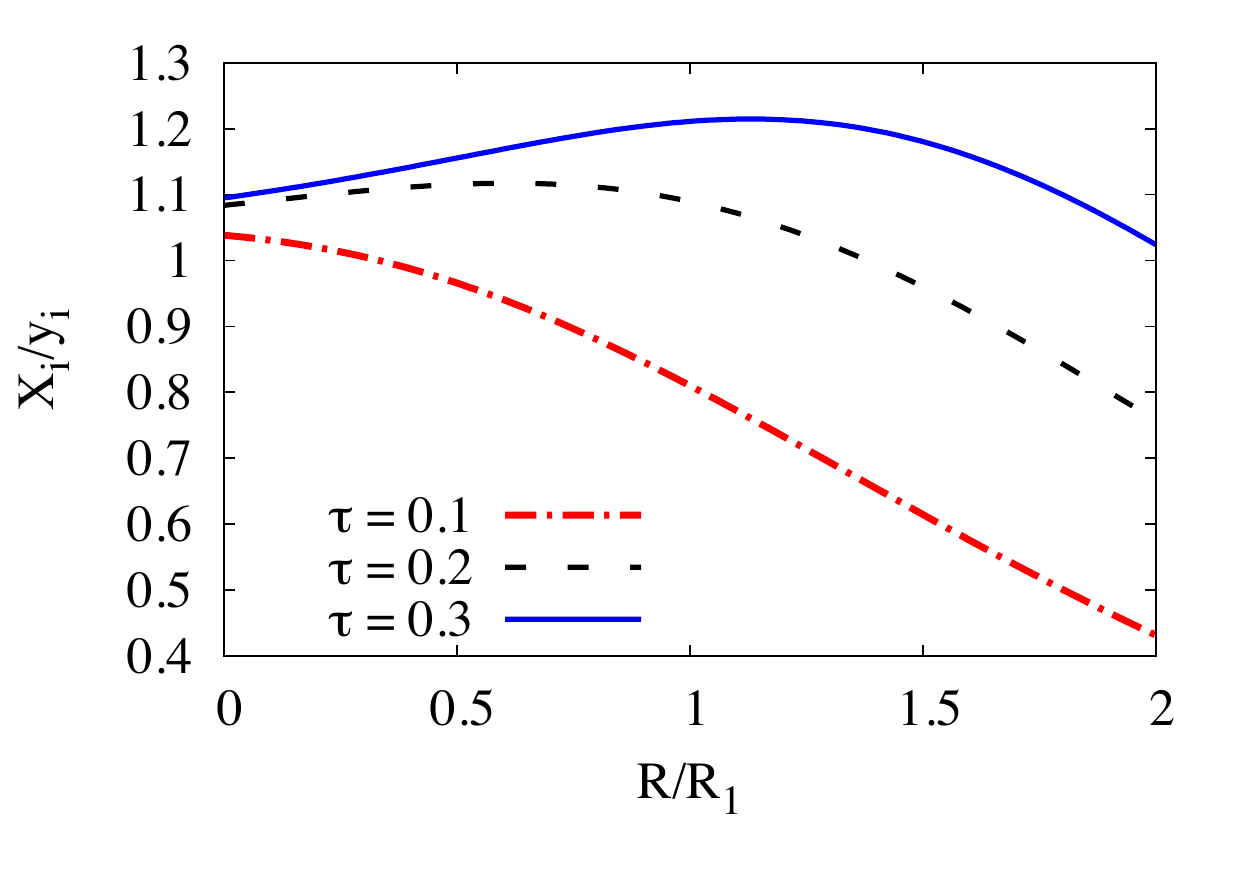}
\caption{Evolution of abundance profiles, in dimensionless units, for an exponential disc forming stars according to the Kennicutt-Schmidt law. The scaling for the Milky Way is $R_1 = 8.75 \; \textrm{kpc}$ and $\tau = 0.27$. Progressive flattening and even gradient inversion occur at late times, contrary to observations. No radial flows, nor inside-out formation, are included in this model.}\label{fig::Conundrum}
\end{figure}

In Fig. \ref{fig::Conundrum} we show, for the fiducial value $q_0 = 37.5$, the resulting profile, in dimensionless units, for some values of the dimensionless time $\tau$.
In the units of the fiducial model, the maximum radius shown is $R_\textrm{max} = 17.5 \; \textrm{kpc}$, while the present time is $\tau = 0.27$, corresponding to a profile between the solid blue line and the dashed black line.

As a consequence of the exponential structure of the disc and of the Kennicutt-Schmidt law, chemical enrichment in this model proceeds on longer time-scales at larger radii (formally, $\sigma$ increases with $\tau$ more rapidly for smaller values of $\rho$, see equation \eqref{sigmaspecialcase}). 
As shown in Fig. \ref{fig::Conundrum}, the inner regions relatively quickly tend towards an approximately common equilibrium, while enrichment at the periphery proceeds much slower, with the outermost annuli being progressively less evolved and more metal-poor. As a result, this class of models predicts, at early times, a negative abundance gradient throughout the disc and, at late times, a progressive flattening in the inner regions and even the development of an inverted gradient there. This prediction is at strong variance with the observed properties of spiral galaxies today (e.g. \citealt{Moustakas+10}; \citealt{Sanchez+14}) and suggests that a major revision of one or more of our hypotheses is necessary.

The most arbitrary of our assumptions is probably equation \eqref{expSFH}. However, we repeated our calculations with  a very different star formation history (vanishing, rather than peaking, for $t = 0$) and we verified that this does not significantly alter the behaviour shown in Fig. \ref{fig::Conundrum}, which is more likely driven by the radial structure of the disc and of the accreting material.

More effective, in this sense, is to remove the hypothesis, implicit in Sec. \ref{sec::ExpKennicuttDisc}, that the stellar scalelength $R_\star$ is constant with time. This route is tightly linked to the inside-out formation scenario (e.g. \citealt{Larson76}), which predicts a rather different distribution of accretion (and therefore of metal dilution) and is indeed the most commonly invoked mechanism to explain abundance gradients in spiral galaxies (e.g.\ \citealt{MF89}; \citealt{MFD97}; \citealt{BP99}; \citealt{Chiappini+01}; \citealt{NO06}; \citealt{Pilkington+12}). The other way to significantly alter the accretion profile is the removal of the independent-annuli assumption (see Sec. \ref{sec::Introduction}). This is the main topic of this work and will be discussed in detail in Sections \ref{sec::AngularMomentumAccretion} and \ref{sec::ChemicalEvolutionRadialFlows}. Our combined approach to inside-out growth and radial flows will be discussed in Section \ref{sec::InsideOutModel}.

\section{Angular momentum, accretion and radial flows}\label{sec::AngularMomentumAccretion}

\subsection{General context and basic equations}\label{sec::BasicEquations}
The classical theory of galaxy formation (\citealt{WR78}) predicts that spiral galaxies accrete a substantial part of their mass from the cooling of large and hot gas reservoirs (\emph{coronae}), which have been detected in X-ray emission around local spiral galaxies, including the Milky Way (\citealt{AndersonBregman2013}; \citealt{Bogdan+13a}; \citealt{Bogdan+13b}; \citealt{MillerBregman2015}). Due to their large (close to virial) temperatures, these structures likely have a non-negligible pressure support against gravity and therefore they are expected to rotate at less than the centrifugal speed at each point; this results in a local angular momentum deficit at the moment of accretion on the centrifugally-supported disc. Other channels, like accretion from filaments (\citealt{DB06}) or minor mergers (\citealt{DiTeodoro}) can bring some cold gas directly to the disc; material  accreted in these ways is not even expected to always rotate in the same sense of the galaxy. An average local angular momentum deficit in accretion is therefore a quite general expectation.

The basic equation describing the dynamical consequences of accretion on to a rotating disc has been first introduced, in the context of galaxy evolution, by \cite{MV81} and then improved by \cite{LF85}. For our purposes, it can be conveniently rewritten, in terms of the radial mass flux $\mu$ defined in \eqref{defmu}, as:
\begin{equation}\label{MVeq}
\mu = -2 \pi \alpha R^2 \dot{\Sigma}_\textrm{acc}
\end{equation}
where we introduced the dimensionless parameter $\alpha$, which is a measure of the local angular momentum deficit of the infalling material with respect to the disc:
\begin{equation}\label{defalpha}
\alpha \vcentcolon= \frac{l_\textrm{disc} - l_\textrm{acc}}{R\frac{\partial l_\textrm{disc}}{\partial R}}
\end{equation}

Since $\partial l_\textrm{disc}/\partial R$ is always positive, the parameter $\alpha$ is positive whenever $l_\textrm{acc} < l_\textrm{disc}$ and the radial mass flux is therefore directed inwards. In the typical case of a flat rotation curve, equation \eqref{defalpha} reduces to:
\begin{equation}\label{alphaMestel}
\alpha = 1 -  \frac{V_\textrm{acc}}{V_\textrm{disc}}
\end{equation}

The simplest situation in which a prediction for $\alpha$ can be made on physical grounds is the case of accretion from an isothermal corona in a logarithmic potential:
\begin{equation}\label{HotMode_alpha}
\alpha = 1 - \sqrt{1 - \delta \frac{c_s^2}{V_\textrm{disc}^2}}
\end{equation}
where $c_s$ is the isothermal sound speed of the hot gas and $\delta$ is the logarithmic derivative of the equatorial density of the corona:
\begin{equation}
\delta = - \frac{\partial \ln \rho}{\partial \ln R}
\end{equation}
Even in this ideal case, $\alpha$ is constant with $R$ only if the coronal density follows a power-law along the equator, which in general will not be true. Furthermore, the shape of the function $\alpha$ can be altered by hydrodynamical interactions between the lower layer of the hot corona and the cold extraplanar gas that circulates near the disc in the so-called \emph{galactic fountain} (\citealt{Marinacci+10}; \citealt{Marasco+12}). We discuss this specific problem in some more detail in the Appendix. In general, however, $\alpha$ is likely a non-trivial function of both space and time; while more theoretical and observational efforts are still required to deepen our understanding in this respect, it is useful to develop a method to evaluate the consequences of equation \eqref{MVeq} on galaxy evolution under general conditions.

\subsection{The general mass flux decomposition}\label{sec::MassFluxDec}
As already mentioned (Sec. \ref{sec::Introduction}, Sec. \ref{sec::EffectiveAccretion}), the observed structural properties of spiral galaxies do not give us direct constraints on the accretion rate surface density $\dot{\Sigma}_\textrm{acc}$, but just on the effective accretion rate $\dot{\Sigma}_\textrm{eff}$, which should then be decomposed in two contributions: direct accretion from IGM $\dot{\Sigma}_\textrm{acc}$ and a radial mass flux $\mu$ (the \emph{mass flux decomposition} problem).

If the equation of angular momentum conservation \eqref{MVeq} is taken into account, the continuity equation \eqref{continuity} becomes a linear differential equation for the unknown $\mu$, which solution is:
\begin{equation}\label{GeneralSolution}
\mu (t, R) = \frac{1}{h(t, R)} \left( \mu_0 - 2 \pi \int_{R_0}^R R' h(t, R') \dot{\Sigma}_\textrm{eff}(t, R') dR' \right) \\
\end{equation}
where $h$ is the dimensionless auxiliary function:
\begin{equation}\label{hdef}
h(t, R) = \exp \left\{ \int_{R_0}^R \frac{dR'}{R' \alpha(t, R')} \right\}
\end{equation}
while $R_0$ is an arbitrary radius and $\mu_0 = \mu(R_0)$ is an integration constant. We recall that $\dot{\Sigma}_\textrm{eff}$ is specified by the considered structural evolution model, according to equation \eqref{dotSigmaEffdef}, while $\alpha$ describes the angular momentum of the accreting material, according to equation \eqref{defalpha}. Once the radial mass flux $\mu$ is known from equation \eqref{GeneralSolution}, the accretion rate surface density $\dot{\Sigma}_\textrm{acc}$ is readily computed from equation \eqref{MVeq}:
\begin{equation}
\dot{\Sigma}_\textrm{acc} = - \frac{\mu}{2 \pi \alpha R^2}
\end{equation}
while the radial velocity $u_R$ immediately comes from equation \eqref{defmu}:
\begin{equation}\label{uRtrivial}
u_R = \frac{\mu}{2 \pi R \Sigma_\textrm{g}}
\end{equation}
This completes the solution of the mass flux decomposition problem.

Note that, imposing the condition that the total accretion rate is finite, it follows form \eqref{MVeq} that the radial mass flux must vanish both at the origin and at infinity, provided that $\alpha$ is a limited function there; in particular, in all but pathological cases, it is safe to assume $R_0 = 0$ and $\mu_0 = 0$ \footnote{Note that, if $\alpha$ is limited near the origin, the definition of $h$ \eqref{hdef} is formally singular for $R_0 = 0$; however, notice also that $R_0$ can be kept arbitrary (and finite) here, since the dependence on it cancels out in equation \eqref{GeneralSolution} for $\mu_0 = 0$.}. We can also see as a consequence, by integrating the continuity equation \eqref{continuity}, that the total accretion rate $\dot{M}_\textrm{acc} = \int_0^\infty 2 \pi R \dot{\Sigma}_\textrm{acc} dR$ always equals the total effective accretion rate. In other words, taking the detailed angular momentum distribution of the accreting material into account does not alter the total needed amount of accretion from the IGM, but it affects the way such accretion is distributed in space.

The solution given by equations \eqref{GeneralSolution} to \eqref{hdef} can be applied to all galaxy models where the net (azimuthally averaged) radial gas flows are dominated by accretion and angular momentum conservation; it provides the mass flux decomposition (and in particular the correct accretion profile) for any desired structural evolution (encoded in $\dot{\Sigma}_\textrm{eff}$) and dynamical properties of accreting material (encoded in $\alpha$), by the means of explicit quadrature formulae.

\subsection{Particular cases}
A particularly simple case is the one where $\alpha$ is a function of time only. This encompasses the majority of the cases considered so far in the literature, which effectively assume $\alpha$ to be constant with both time and space. If $\alpha$ does not dependent of radius, the solution \eqref{GeneralSolution}, \eqref{hdef} reads:
\begin{equation}\label{constalphasol}
(-\mu)(t, R) = 2 \pi R^{- \frac{1}{\alpha(t)}} \int_0^R R'^{1 + \frac{1}{\alpha(t)}} \dot{\Sigma}_\textrm{eff}(t, R') dR'
\end{equation}

An interesting property of equation \eqref{constalphasol} is that, whenever the effective accretion rate $\dot{\Sigma}_\textrm{eff}$ is (or can be approximated by) an analytic function of radius:
\begin{equation}
\dot{\Sigma}_\textrm{eff} = \sum_k a_k R^k
\end{equation}
then the derived accretion rate $\dot{\Sigma}_\textrm{acc}$ is an analytic function of radius as well:
\begin{equation}
\dot{\Sigma}_\textrm{acc} = \sum_k \frac{a_k}{1+(k+2)\alpha}R^k
\end{equation}
For the simple exponential model described in Sec. \ref{sec::ExpKennicuttDisc}, equation \eqref{constalphasol} becomes:
\begin{equation}\label{musol}
(-\mu)(t, R) = \dot{M}_\star(t) f\left(\frac{1}{\alpha(t)}, \frac{R}{R_\star}\right) + \dot{M}_\textrm{g}(t) f\left(\frac{1}{\alpha(t)}, \frac{R}{R_\textrm{g}}\right)
\end{equation}
where $f$ is derived from the lower incomplete Euler gamma function $\gamma$:
\begin{equation}
f(a, x) \vcentcolon= x^{-a} \gamma (2+a, x) = \sum_{\substack{k = 0}}^\infty \frac{(-1)^k}{k!}\frac{x^{k+2}}{a+k+2}
\end{equation}
while the accretion rate surface density is given by:
\begin{equation}\label{dotSigmaAccExp}
\dot{\Sigma}_\textrm{acc}(t, R) = \frac{\dot{M}_\star(t)}{2 \pi R_\star^2}g\left(\frac{1}{\alpha(t)}, \frac{R}{R_\star}\right) + \frac{\dot{M}_\textrm{g}(t)}{2 \pi R_\textrm{g}^2}g\left(\frac{1}{\alpha(t)}, \frac{R}{R_\textrm{g}}\right)
\end{equation}
with:
\begin{equation}
g(a, x) \vcentcolon= a x^{-(2+a)}\gamma(2+a, x) = \sum_{\substack{k = 0}}^\infty \frac{(-1)^k}{k!}\frac{a}{a + k + 2} x^k
\end{equation}

The expressions above further simplify for some particular integer values of $a$, or particular fractional values of $\alpha$. Completely obvious is the case $\alpha = 0$ (i.e. $a \to \pm \infty$), where it is $\mu = 0$ and $\dot{\Sigma}_\textrm{acc} = \dot{\Sigma}_\textrm{eff}$, as expected. Less trivial particular cases are essentially those originally found and described by \cite{PT89} \footnote{Some differences arise not from the decomposition, but from the details or their assumed $\dot{\Sigma}_\textrm{eff}$.}. Our approach allows us to explore, with the same simplicity coming from an analytic formulation, the whole continuum of possible values of $\alpha$.

There are of course other cases admitting a simplified treatment. For instance, the one where $\alpha$ is a linear function of radius:
\begin{equation}
\alpha(R) = \alpha_0 \left(1 + \frac{R}{R_\alpha}\right)
\end{equation}
corresponds to the following simple form for the auxiliary function:
\begin{equation}
h(R) = K \left( 1 + \frac{R_\alpha}{R} \right)^{-\frac{1}{\alpha_0}}
\end{equation}
where $K$ is a constant such that $h(R_0) = 1$. Note that the coefficients $\alpha_0$ and $R_\alpha$ can be a function of time. Note also that $R_\alpha$ can be negative, but in this case there is a critical radius $|R_\alpha|$, where corotation occurs, out of which a separate solution is formally needed, associated to an outward radial flow.

\subsection{Application to the minimal galaxy evolution model}\label{sec::FluxDecompositionApplication}

As an illustrative example, we discuss here the application of our analytic mass flux decomposition to the fiducial model of Sec. \ref{sec::ExpKennicuttDisc}.

\begin{figure}
\centering
\includegraphics[width=9cm]{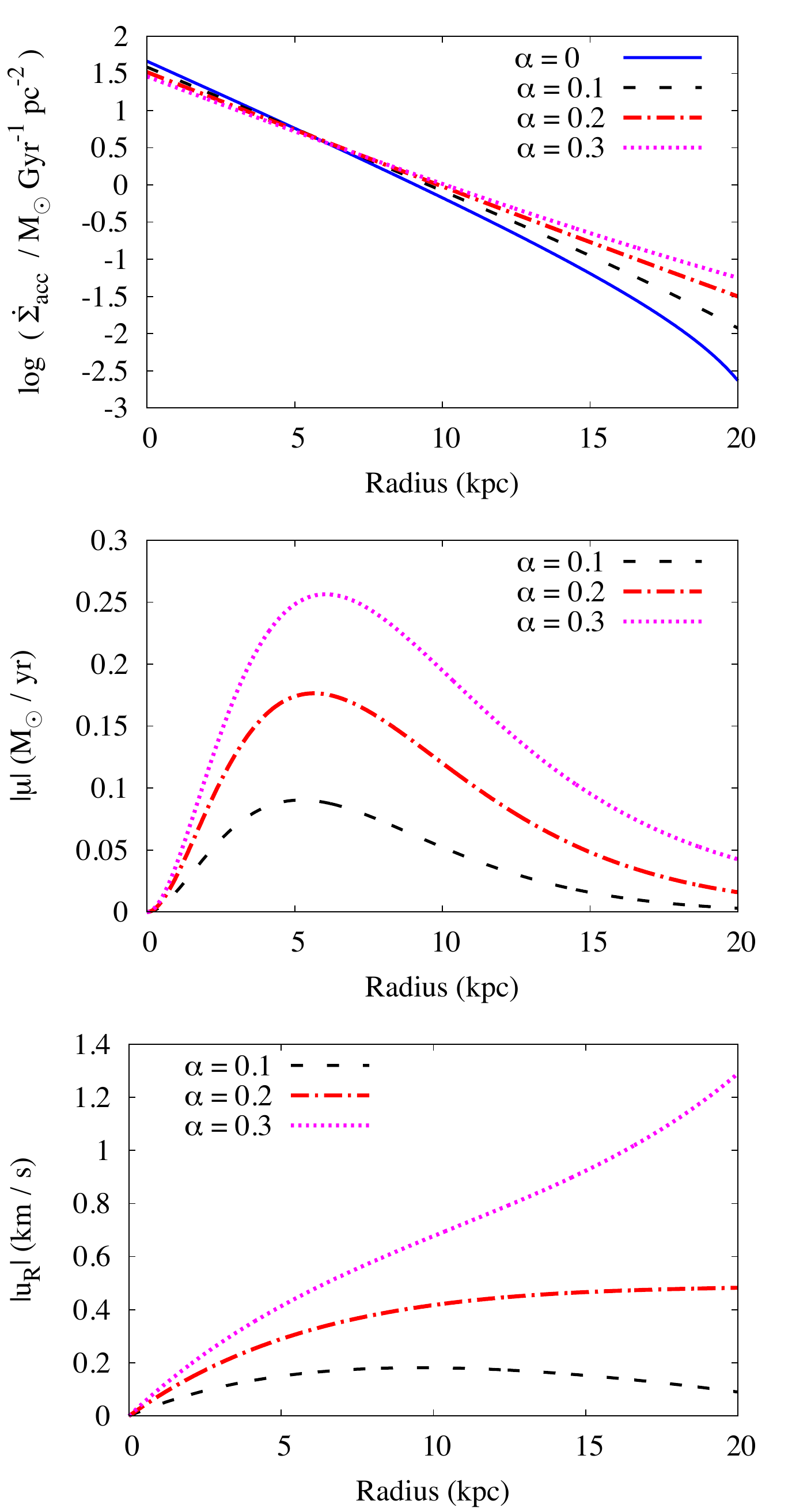}
\caption{Illustrative mass flux decomposition for the fiducial model (Sec. \ref{sec::ExpKennicuttDisc}), at the present time, for different values of the angular momentum mismatch parameter $\alpha$. Upper panel: accretion rate surface density (in logarithmic scale). Middle panel: inward radial mass flux. Lower panel: inward radial velocity. For $\alpha = 0$, there are no radial flows and accretion coincides with the effective accretion. For larger $\alpha$, accretion from the IGM preferentially occurs in the outer regions, from which the gas is then brought inwards by radial flows.}\label{fig::FluxPlot}
\end{figure}

Fig. \ref{fig::FluxPlot} reports the accretion profile (upper panel), the inward radial mass flux $(-\mu)$ (middle panel) and the inward radial velocity $(-u_R)$ (lower panel), computed at the present time for some values of the angular mismatch parameter $\alpha$, assumed to be constant with $R$.

The case $\alpha = 0$ is for fine-tuned local angular momentum balance between accretion and the disc. In this case, there are no radial flows and the accretion rate surface density (solid blue line) is equal to the effective accretion rate surface density \eqref{dotSigmaEffExp}.

Slightly larger values of $\alpha$ already have a significant impact on the shape of the accretion profile. The qualitative effect is a depletion of the needed accretion in the inner regions and a corresponding enhancement in the outskirts. This is a redistribution effect, since the total accretion rate is independent of $\alpha$ (see Sec. \ref{sec::MassFluxDec}). The magnitude of the effect rapidly increases with increasing $\alpha$: already for $\alpha = 0.2$ (corresponding, for a flat rotation curve, to material accreting with 80\% of the rotational velocity of the disc), the needed accretion at 20 kpc is increased by more than one order of magnitude.

The increasing change in the shape of the accretion profile is associated to the onset of an increasingly large inward radial mass flux: the material falling down in the outskirts then travels radially within the disc, to reach the inner regions where it is needed to sustain the structural evolution of the disc. While the overall shape of the radial mass flux profile is mainly dictated by general boundary condition requirements (see Section \ref{sec::MassFluxDec}), the precise position of the peak is in general a function of $\alpha$. However, it always coincides with the boundary between the inner region, where accretion is depleted with respect to the effective accretion, and the outer one, where it is enhanced. This fact is completely general, being an obvious consequence of the continuity equation \eqref{continuity}.

The predicted radial velocity pattern is a non-trivial function of both radius and $\alpha$ (besides, of course, of time, as all the involved quantities), implying that the full calculation is always required for a model with radial flows to be compatible with angular momentum conservation. The predicted magnitude of radial velocities is very low, of the order of 1 $\textrm{km}\; \textrm{s}^{-1}$ or less, which is not directly accessible to observations, but important for chemical evolution (see Sec. \ref{sec::Introduction} and the next Section).

All the trends reported in Fig. \ref{fig::FluxPlot} are in very good agreement with the ones found by \cite{BS12} by means of numerical techniques.

\section{Chemical evolution with radial flows}\label{sec::ChemicalEvolutionRadialFlows}
In the presence of radial flows, equation \eqref{metaleq0} is no longer valid and it has to be replaced with the following equation:
\begin{equation}\label{metaleq1}
\frac{\partial \tilde{X}_i}{\partial t} + u_R\frac{\partial \tilde{X}_i}{\partial R} = \frac{\dot{\Sigma}_\star}{\Sigma_\textrm{g}} - \tilde{X}_i\frac{\dot{\Sigma}_\textrm{acc}}{\Sigma_\textrm{g}}
\end{equation}

This is a partial differential equation, the solution of which is usually approached by means of numerical finite-difference techniques.
Equation \eqref{metaleq1}, however, has the special property of being linear in the unknown $\tilde{X}_i$. This allows it to be solved in a much simpler, more stable and less numerically demanding way, which is the method of characteristics.

\subsection{The method of characteristics}
\begin{figure*}
\centering
\includegraphics[width=18cm]{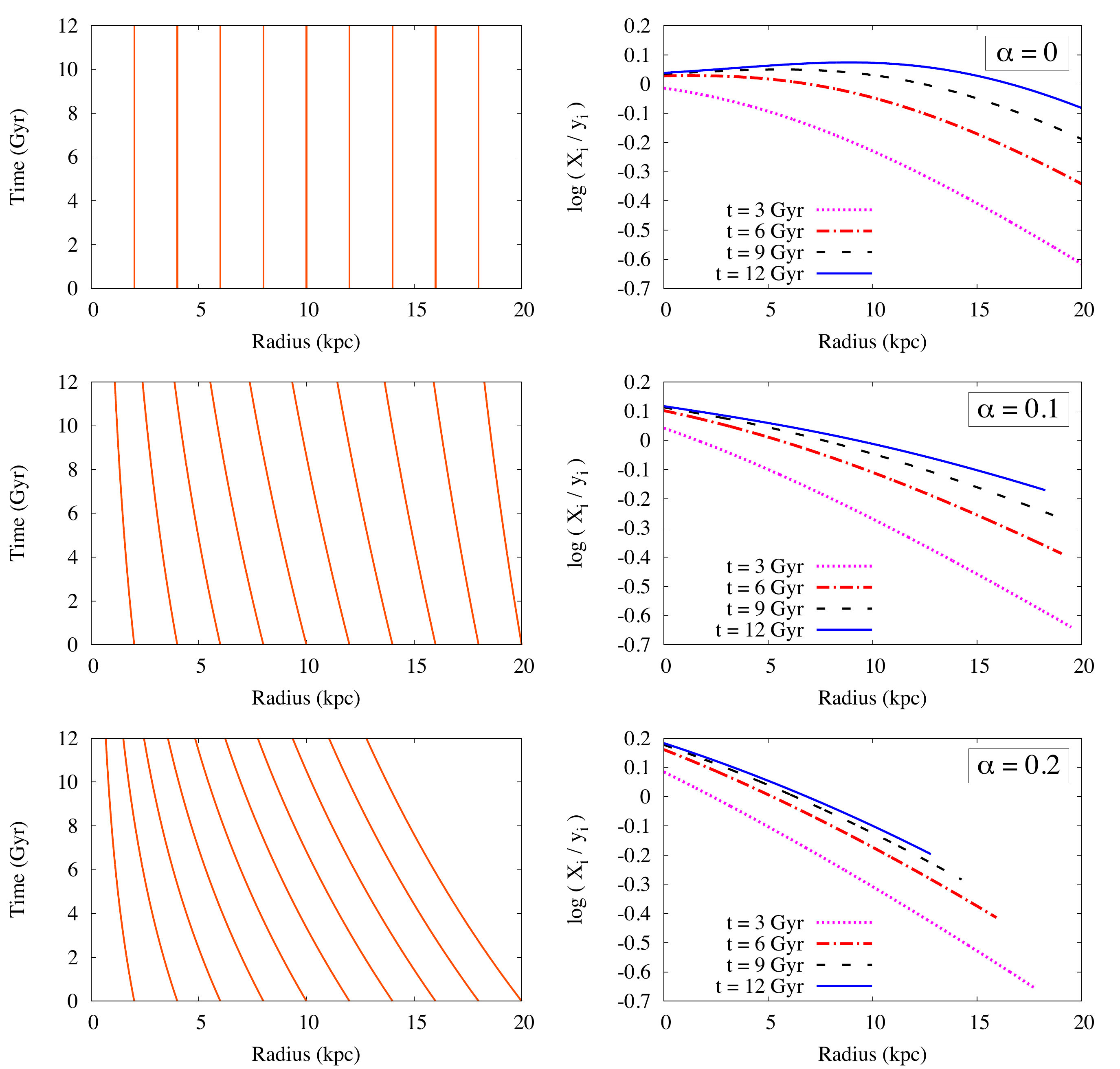}
\caption{Illustrative application of the method of characteristics to the fiducial model of Sec. \ref{sec::ExpKennicuttDisc}, when radial flows are included as in Sec. \ref{sec::FluxDecompositionApplication}. Left-hand panels: the characteristic lines, along which metallicity evolution can be computed independently. Right-hand panels: the resulting logarithmic abundance profiles, at different epochs; the solid blue line is for the present time. Different rows are for different values of the angular momentum mismatch parameter: $\alpha = 0$ (upper panels), $\alpha = 0.1$ (middle panels) and $\alpha = 0.2$ (lower panels). The case $\alpha = 0$ (no radial flows) is equivalent to the independent-annuli assumption (cfr. Fig. \ref{fig::Conundrum}). Negative gradients are already predicted for small, but non vanishing, values of $\alpha$.}\label{fig::CharPlot}
\end{figure*}

The method of characteristics consists of the reduction of a linear partial differential equation into a combination of two sets of ordinary differential equations. The first set defines characteristic lines, which are the loci, in the computational domain, along which the information propagates that is relevant to the solution of the problem. The second set encodes such an information propagation and describes the evolution of the unknown field along individual characteristic lines.

In our particular case, characteristic lines, in the $(t,R)$ computational plane, are the integral curves of the radial velocity field $u_R$: that is, they are the solutions of the equation:
\begin{equation}\label{chareq}
\frac{dR_\textrm{char}}{dt} = u_R \\
\end{equation}

The left-hand side of equation \eqref{metaleq1} is easily recognized as the total time derivative of $\tilde{X}_i$ along the characteristic lines defined by equation \eqref{chareq}. This implies that, when restricted along characteristics, equation \eqref{metaleq1} formally reduces to equation \eqref{metaleq0} and therefore that the metallicity evolution along each characteristic can be independently computed by means of the explicit quadrature formulae \eqref{metalsol} and \eqref{defsigma}.

In this way, the problem of including radial gas flows into chemical evolution is reduced to replacing a model with independent annuli with a model with independent characteristics, essentially keeping the same computational difficulty. Note, however, that the independence of characteristic lines strictly relies on our simplifying assumptions and in particular on the instantaneous recycling approximation (Sec. \ref{sec::IndependentAnnuliMetalEvol}). For our approach to be extended to elements produced by long-lived stars (among all, iron), some interaction should be accounted for between different lines, within distances given by $|u_R|$ times the relevant stellar lifetimes; furthermore, the distance travelled by the stars themselves, due to stellar radial migration, should be taken into account as well in this case.

From the mathematical point of view, our approach is similar to the one by \cite{EdmundsGreenhow95}, although these authors did not include any accretion term in their equations, neither as a dynamical driver of radial flows, nor as a source for dilution of metals, while our formalism includes both effects in a natural way.

Physically, characteristics may be regarded as shrinking gaseous rings, though attention should be paid to the fact that the matter they are constituted of is not fixed but continuously changing with time: at each radius, novel material is acquired from the IGM, while some other mass is left behind, deposited into the stellar component due to star formation.

\subsection{Application to the minimal model}
In Fig.\ \ref{fig::CharPlot}, we illustrate the application of the method of characteristics to the chemical evolution of the minimal model of Sec. \ref{sec::ExpKennicuttDisc}, with radial flows induced by a local angular momentum mismatch, as described in Sec. \ref{sec::FluxDecompositionApplication}, assuming some values of $\alpha$, which we take here to be constant with both space and time, increasing from $\alpha = 0$ (upper panels) to $\alpha = 0.2$ (lower panels). In the left-hand panels, the characteristic lines are drawn in the $(t, R)$ plane, while the right-hand panels show the computed logarithmic abundance profiles, as a function of radius $R$, for different times.

For $\alpha = 0$, there are no radial flows ($u_R = 0$ everywhere) and therefore, according to equation \eqref{chareq}, characteristic lines are lines of constant radius. This case is, of course, coincident with the independent annuli model. In fact, the abundance profile shown in the upper right panel is just the same as Fig. \ref{fig::Conundrum}, but with dimensional units for space and time and a logarithmic scale for the abundances. As already pointed out, the predicted abundance profile at the present time (the solid blue line) is in very strong disagreement with observations.

For $\alpha = 0.1$ and $\alpha = 0.2$, small, but non-negligible, inward radial gas flows onset (cfr. Fig. \ref{fig::FluxPlot}, lower panel). Accordingly, characteristic lines now connect radii that are no more constant, but decreasing with time. From the right-hand panels, we see that even a small deviation from perfect corotation between the accreting material and the disc has a dramatic impact on the predicted abundance profiles. A very small value of the angular momentum mismatch parameter $\alpha = 0.1$ is already able to completely remove the inner gradient inversion, which was plaguing the independent-annuli model. Furthermore, the steepness of the resulting profile is strongly dependent on $\alpha$. 
\subsection{Origin of the steepening effect}
Both the mathematical and the physical origin of the described behaviour can be relatively easily understood (see also \citealt{ChamchamTayler94} on this).

From a mathematical point of view, the origin is two-fold. First, as described in Sec. \ref{sec::FluxDecompositionApplication} (Fig. \ref{fig::FluxPlot}, upper panel) increasing $\alpha$ implies a modification of the accretion profile. Depleted accretion in the inner regions means a reduced dilution and therefore higher metallicity there, while the opposite is true in the outskirts, going in the direction of creating or steepening abundance gradients. Furthermore, radial flows tend to bring inwards metals produced at large radii, further increasing the enrichment of the inner regions at the expenses of the outer ones.

From a physical point of view, the two effects are tightly linked to each other and they can be understood altogether. In our approach, in fact, all the models share the same structural evolution and therefore the same amount of gas arriving at each radius and time; the only difference is in the path that is followed by gas elements to come to their position: directly from the IGM, and therefore unpolluted, in the case $\alpha = 0$, or through a more complex path, in the other cases, including a travel within the disc, where primordial material gets mixed with higher metallicity gas and polluted by ongoing star formation.

\subsection{The role of boundary conditions}\label{subsubsec::Boundaries}
\begin{figure*}
\centering
\includegraphics[width=18cm]{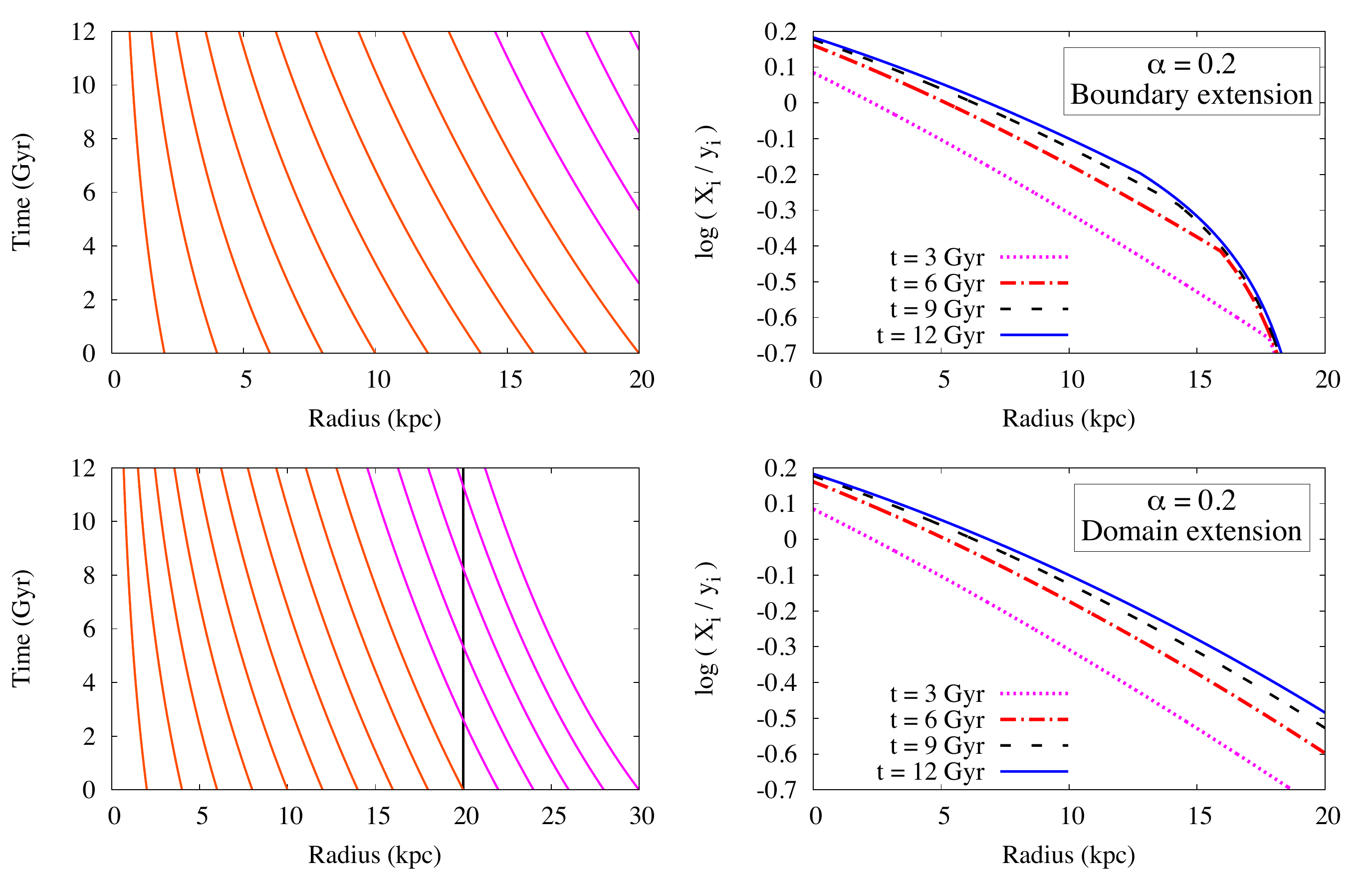}
\caption{Similar to Fig. \ref{fig::CharPlot}, but showing the two alternative strategies to extend the predicted abundance profiles to the whole computational domain, in the illustrative case $\alpha = 0.2$. Upper panels: the boundary-extension strategy, where the $(t, R)$ plane is filled with additional characteristics starting from the maximum radius $R_\textrm{max}$ at different times; this is equivalent to the classical approach and predicts a sharp steepening of gradients at the disc periphery. Lower panels: the domain-extension strategy, where the additional characteristics start at the initial time $t = 0$, but from larger radii; the outer profiles in this case are the smooth continuation of the ones in Fig. \ref{fig::CharPlot}. Note that the characteristics of the boundary-extension strategy are the same of the domain-extension strategy, but cut along the vertical black solid line (lower left panel).}\label{fig::CharPlotBoundaries}
\end{figure*}
A major feature of Fig. \ref{fig::CharPlot} is that, when $\alpha \neq 0$, the domain where abundances are predicted appears to be shrinking with time. Such a behaviour is a logical outcome of our chosen computational method: at each time, the maximum radius where abundances are predicted is the one reached, at that time, by the outermost considered characteristic: in the shown example, the one starting at $t= 0$ from a radius $R_\textrm{max} = 20 \; \textrm{kpc}$. Points lying, in the $(t, R)$ plane, above and to the right of such a line are not reached by any characteristic and therefore the information needed to compute the metallicity evolution is not available there.

A simple solution to this drawback is to fill the gap in the $(t, R)$ plane with additional characteristics, as shown, for the illustrative case $\alpha = 0.2$, in the upper-left panel of Fig. \ref{fig::CharPlotBoundaries}. At variance with the lines already shown in Fig. \ref{fig::CharPlot}, lower left panel, all of which start at the same time $t = 0$ but from different radii, these additional lines all start from the same radius $R = R_\textrm{max}$ but at different times. This method, which we refer to as the \emph{boundary-extension} strategy, allows the full calculation of the metallicity evolution, provided that some value is chosen for the normalized abundance $\tilde{X}_i$ at the starting point of the added lines, which coincides with the outermost considered radius $R_\textrm{max}$. This strategy is the equivalent, within the characteristic framework, of the one adopted in classical finite-difference calculations, where some boundary condition needs to be stated for the metallicity of the material incoming from the chosen outer edge of the model.

The upper-right panel of Fig. \ref{fig::CharPlotBoundaries} shows the predictions  of the boundary-extension strategy, if accretion of unpolluted material is allowed to occur  at $R = R_\textrm{max}$. Profiles computed in this way show a sharp steepening in the outer regions. The illustrative example shown here is an extreme case, the effect being milder for a non-primordial composition of the IGM. For instance, if we assumed $\tilde{X}_{i, \textrm{IGM}} = 0.1$, all predicted abundances would be slightly higher and the profiles in the upper-right panel of Fig. \ref{fig::CharPlotBoundaries} would tend, for $R \to R_\textrm{max}$, to $\log(\tilde{X}_i) = -1.0$ instead of $\log(\tilde{X}_i) = -\infty$.

The solution discussed above is not the only possibility. The precise location and extent of the information gap in the $(t, R)$ plane clearly depends on the choice of the initial radius for the outermost characteristic $R_\textrm{max}$. Since this is in general largely arbitrary, it is useful to look at the results for different choice of $R_\textrm{max}$. In particular, a suitably large value of $R_\textrm{max}$ should in principle allow to fill any desired region of the $(t, R)$ plane in a natural way. This is shown in the lower left panel of Fig. \ref{fig::CharPlotBoundaries}, where characteristics start from radii as large as $R_\textrm{max} = 30 \; \textrm{kpc}$ and allow predictions out to $R = 20 \; \textrm{kpc}$ at $t = 12 \; \textrm{Gyr}$, without the need of particular assumptions on boundary condition \footnote{The reader may remember from Sec. \ref{sec::Conundrum} that our fiducial model without radial flows formally develops a wind outside 20 kpc. When radial flows are considered, however, the accretion profile remains positive out to large radii already for small values of $\alpha$ and can therefore be safely used in equations \eqref{metaleq0} and \eqref{defsigma}.}. 

The abundance profiles predicted by this \emph{domain-extension} strategy are reported in the lower right panel of Fig. \ref{fig::CharPlotBoundaries}. In this case, the predicted abundance profiles are a smooth continuation of the ones in Fig. \ref{fig::CharPlot} (lower right panel) and no steepening is predicted in the outer regions.

The difference between the predictions of the two methods, which may be thought of as the extremes of a continuum of intermediate possibilities, can be very clearly understood by looking at the corresponding characteristic line diagrams. Comparing the two left-hand panels of Fig. \ref{fig::CharPlotBoundaries}, in fact, we see that the set of characteristics used in the boundary-extension method are the same of the domain-extension one, but cut along the line of constant radius $R = 20 \; \textrm{kpc}$. This has the consequence of ignoring the initial metallicity evolution, occurring along the cut portion of the lines, and therefore of underestimating the metallicity in the outer regions.

We cannot say that one method is definitely better than the other, since we do not know, in general, whether and how much radial accretion, in real galaxies, directly occurs from an outer boundary. However, irrespectively of the adopted choice, it should be emphasized that the method of characteristics is the simplest way to clearly identify the region where predictions on abundance profiles, in a given galaxy evolution model, are dominated by effects associated with the choice of boundary conditions.
This is potentially of great importance in the modelling of the outskirts of spiral galaxies, where theoretical predictions often disagree with each other and with observations.

\subsection{Calibration from observed gradients}\label{sec::Calibration}
For a quantitative comparison with observations, we choose the domain extension technique, so that predictions do not depend on the choice of $R_\textrm{max}$, provided that it is large enough. In our application, we never found an extension to be necessary farther out than $40 \; \textrm{kpc}$.
\begin{figure}
\centering
\includegraphics[width=9cm]{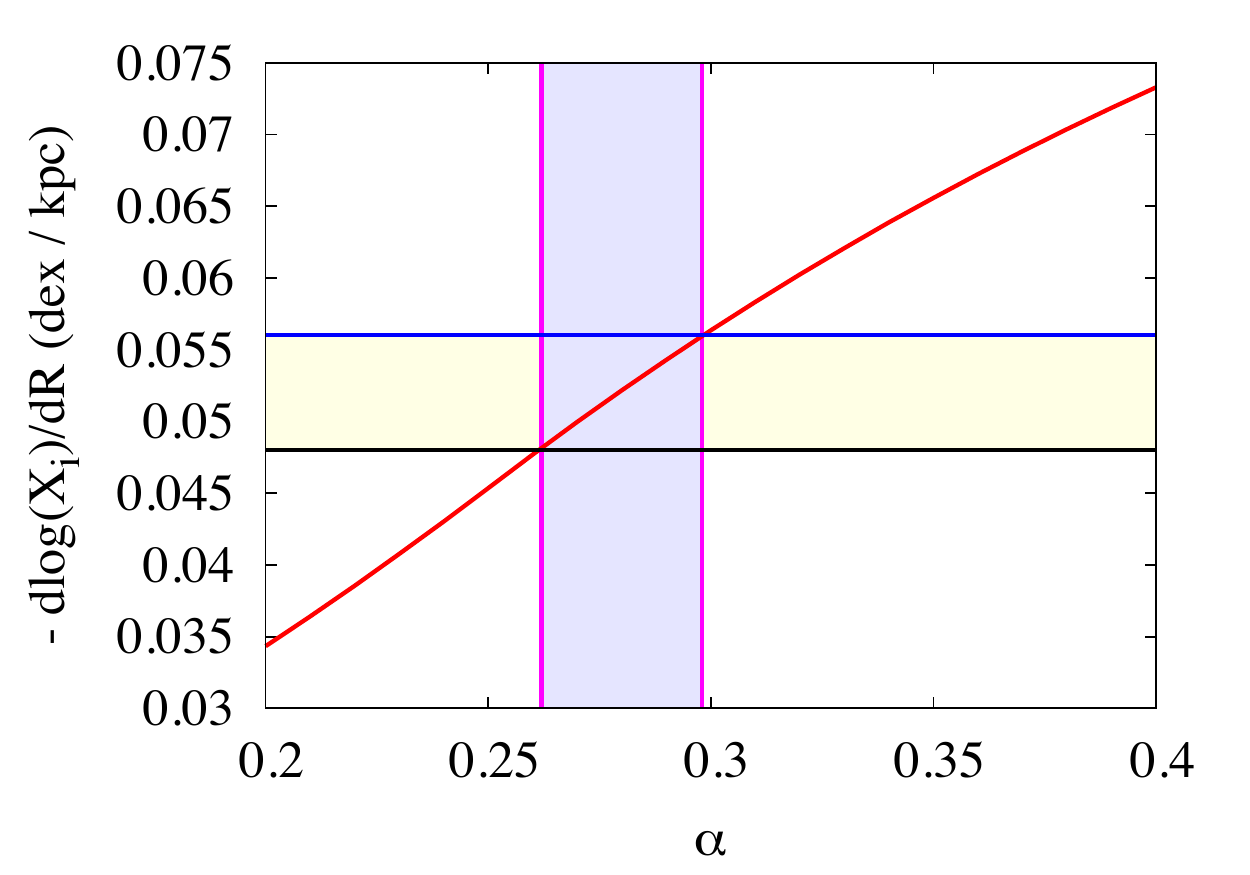}
\caption{Abundance gradients, in the fiducial model, as a function of the angular momentum mismatch parameter $\alpha$. Horizontal lines are Milky Way measurements for oxygen (upper line) and magnesium (lower line) in Cepheids (\citealt{LL11}). The predicted gradient (solid red line) is very sensitive to the rotational velocity of accreting gas; observations constrain it in a quite narrow range, of about $72 \pm 2$ per cent of the rotational velocity of the disc.}\label{fig::GradPlot}
\end{figure}

In Fig. \ref{fig::GradPlot} the predictions are shown of our fiducial model for the present-day abundance gradient, computed between 5 and 17 kpc, as a function of the angular momentum mismatch parameter $\alpha$ and they are compared with the observed gradient, measured in the same radial range by \cite{LL11} for two representative $\alpha$ elements, oxygen and magnesium, in Cepheid stars in the Milky Way. Being very young stars, Cepheids are a fair tracer of ISM abundances, while oxygen and magnesium are among those elements that are predominantly produced by Type II supernovae (SN) and therefore they are negligibly affected by time-delay effects, which are not included in our calculations. Besides observational uncertainties, the difference between the observed gradients for oxygen and magnesium may be partially due to a different dependence of the stellar yields on metallicity. Since our formalism does not take these details into account, our predictions can just be compared with the finite-width range defined by the two observed values. However, the predicted gradient is so sensitive to the angular momentum of accreting material that $\alpha$ is nonetheless constrained in a very narrow range, between 0.26 and 0.30. In the approximation of a flat rotation curve, (equation \eqref{alphaMestel}), this means that metal-poor gas is accreted on to the Milky Way with an average rotational velocity equal to $70 - 74$ per cent the rotational velocity of the disc, in very good agreement with the estimate by \cite{BS12}. Inverting \eqref{HotMode_alpha}, this would correspond to a coronal temperature $T_\textrm{corona} \sim 1.8/\delta \times 10^6 \; \textrm{K}$, assuming a circular speed $V_\textrm{disc} = 220 \; \textrm{km}/\textrm{s}$ and a mean molecular weight $\mu_\textrm{m} = 0.6$.

\section{Inside-out models with radial flows}\label{sec::InsideOutModel}
\subsection{Structural constraints to inside-out growth}\label{sec::InsideOutParameters}

Radial flows are not the only possibility to explain abundance gradients in spiral galaxies, the main alternative being inside-out growth (see references in Sec. \ref{sec::Conundrum}). 

The angular momentum build-up of galaxies, predicted by the theory of tidal torques (\citealt{Peebles1969}; \citealt{White1984}) is believed to be a major ingredient in determining the structure of galaxy discs (\citealt{FallEfstathiou1980}; \citealt{MoMaoWhite1998}), which are generally expected to grow in size with time (e.g.\ \citealt{FA09}; \citealt{Brooks+11}), though the details depend on the particular accretion and merger histories of individual galaxies (\citealt{AWN2014}). On the other hand, the angular momentum assembly of spiral galaxies is also expected to drive radial flows within the disc, if the local aspects of angular momentum accretion are taken into account (Sec. \ref{sec::BasicEquations}). It is therefore likely that inside-out growth and radial flows, rather than being alternative, are both at work in real galaxies. However, the problem arises of how to break the degeneracies of a combined approach and to tell what contribution to the development of abundance gradients comes from inside-out growth and radial flows separately.

In close resemblance to what we have done in Sec. \ref{sec::MinimalModel} for a simpler model, we propose to constrain the structural evolution (in this case, the inside-out growth) by using observed structural properties (e.g. the present-day distribution of gas, stars and star formation rate). In particular, the radial profile of the star formation rate surface density (SFRD) has been shown to be a very good tracer of inside-out growth (e.g. \citealt{MM+07}; \citealt{Pezzulli+15}). We limit here to discuss the illustrative case of the Milky Way, leaving the application to external galaxies for a future work. 

We adopt here a formalism inspired by the classical semi-analytic models of inside-out growth (e.g.\ \citealt{MF89}; \citealt{BP99}, \citealt{Chiappini+01}). We assume the effective accretion rate surface density to be an exponentially decreasing function of time, but with a time-scale that is an increasing function of radius:

\begin{equation}\label{ClassicalInfallLaw}
\dot{\Sigma}_\textrm{eff}(t, R) = \frac{\Sigma_\infty(R)}{t_\textrm{acc}(R)} \exp\left(-\frac{t}{t_\textrm{acc}(R)}\right)
\end{equation}
where $\Sigma_\infty$ is the asymptotic total (baryonic) mass surface density of the disc and $t_\textrm{acc}$ is the radially dependent accretion time-scale. To parametrize $\Sigma_\infty$, we require the total baryonic mass of the disc to have an exponential radial distribution at late times and therefore:
\begin{equation}
\Sigma_\infty(R) = \frac{M_\infty}{2 \pi R_\infty^2} e^{-\frac{R}{R_\infty}}
\end{equation}
while for the accretion time-scale $t_\textrm{acc}$ we choose, as usual, a linearly increasing function of radius:
\begin{equation}
t_\textrm{acc}(R) = t_{\textrm{acc}, 0}\left(1 + \frac{R}{R_\textrm{acc}}\right)
\end{equation}
This model has 4 parameters $(M_\infty, R_\infty, t_{\textrm{acc}, 0}, R_\textrm{acc})$, which we constrain making a fit on observed properties of the disc today.

Our main observational constraint is the present-day SFRD profile of the Milky Way, as traced by the distribution of SN remnants (\citealt{Case+98}). Since this is only given in dimensionless units, we scale it to the solar value, which we estimate applying the Kennicutt-Schmidt law \eqref{Klaw} to a gas surface density of $10 \; \textrm{M}_\Sun \; \textrm{pc}^{-2}$ (\citealt{BM98}, note that we included a correction for helium). The SFRD profile derived in this way is also consistent with several other estimates of the SFRD in our Galaxy (e. g. \citealt{FT12} and references therein)\footnote{Note, however, that the \cite{Case+98} profile has been recently questioned by \cite{Green2015}, who propose a more centrally concentrated distribution.}. We also require that the mass surface density of stars at the solar radius ($R_\Sun = 8.5 \; \textrm{kpc}$), is $\Sigma_\star(R_\Sun) = (37.1 \pm 1.2) \; \textrm{M}_\Sun \; \textrm{pc}^{-2}$ (\citealt{Read2014}) and that the stellar scalelength today is $R_\star = (2.5 \pm 0.25) \; \textrm{kpc}$. This last constraint is chosen in such a way that the nominal $2 \sigma$ interval coincides with the range of possible values $2-3$ kpc quoted by \cite{BT08}.

\begin{figure}
\centering
\includegraphics[width=9cm]{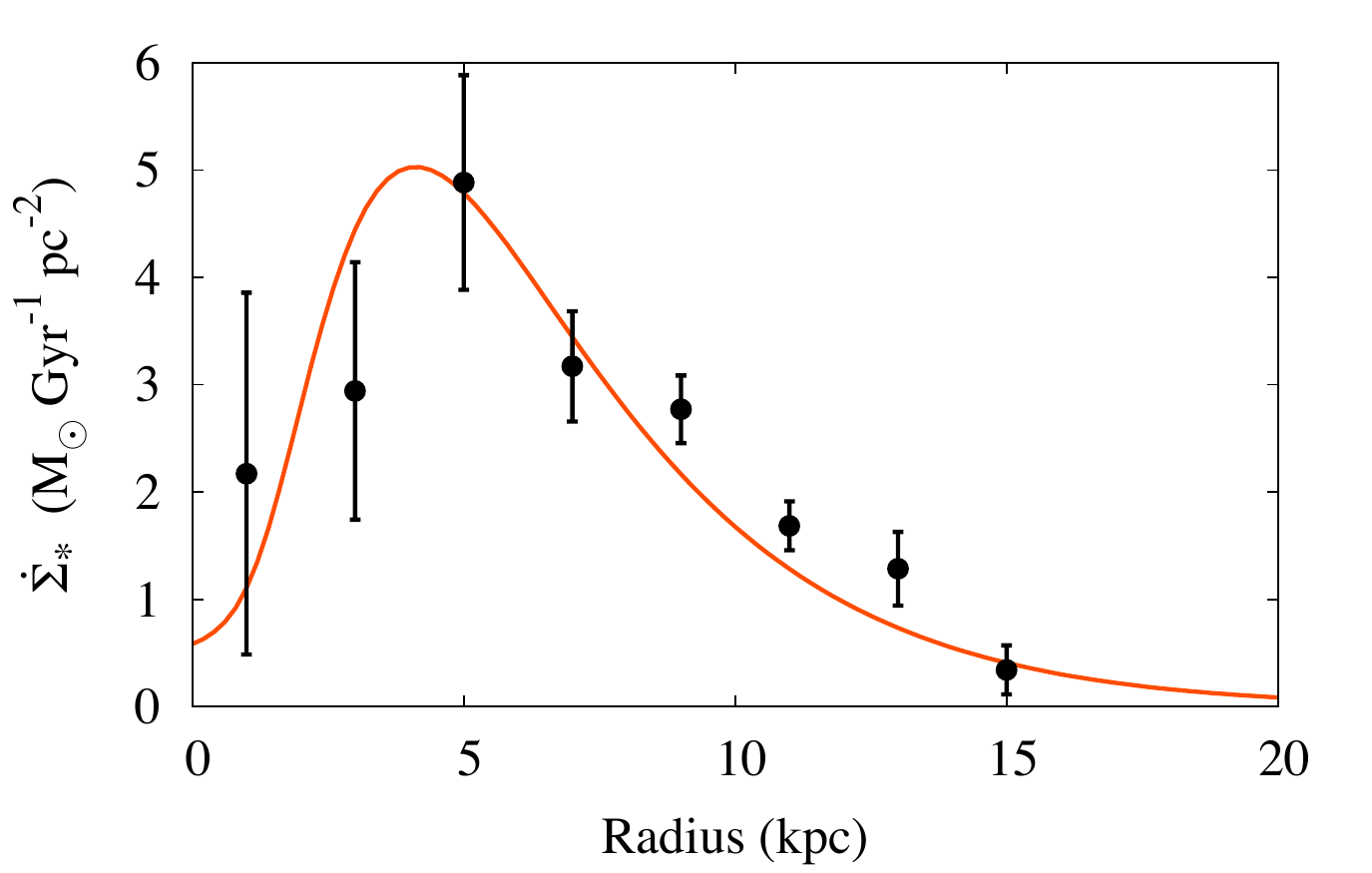}
\caption{Calibration of the inside-out growth parameters on the observed SFRD profile. Points with error bars are measurements for the Milky Way based on the distribution of SN remnants (\citealt{Case+98}). The solid line is our model. See text for further details.}\label{fig::SFRD}
\end{figure}

For each set of parameters, equations \eqref{ClassicalInfallLaw} and \eqref{Klaw} can be integrated in time, yielding the surface density of stars and gas (and thence of star formation rate) at each time and radius. Then, by means of a $\chi^2$ minimization, we select the model that better matches the observational requirements. In Fig. \ref{fig::SFRD} the comparison is shown between the model and observed SFRD profile at the present time. The model present-day stellar mass surface density at the solar radius is $\Sigma_\star(R_\Sun) = 36.0 \; \textrm{M}_\Sun \; \textrm{pc}^{-2}$, while the present-day stellar scalelength is $R_\star = 2.9 \; \textrm{kpc}$. The inferred infall time-scale virtually vanishes at the origin (formally, $t_{\textrm{acc},0} = 22 \; \textrm{Myr}$) and then linearly increases with radius with a slope of $1.1 \; \textrm{Gyr}/\textrm{kpc}$.

Notice that with the formalism adopted here the stellar disc is only approximately exponential at any given time. A treatment where the stellar disc is strictly exponential at all times, with a time increasing scalelength (\citealt{Pezzulli+15}) gives similar results (Pezzulli \& Fraternali, in preparation). Another alternative, adopted by \citet{BS12}, is to assume that the disc of gas (rather than stars) is exponential at all times; in this case the SFRD is exponential as well (because of the Kennicutt-Schmidt law) and in particular there is no depletion of star formation at small radii.

\subsection{Accretion and radial flows in an inside-out model}
\begin{figure}
\centering
\includegraphics[width=9cm]{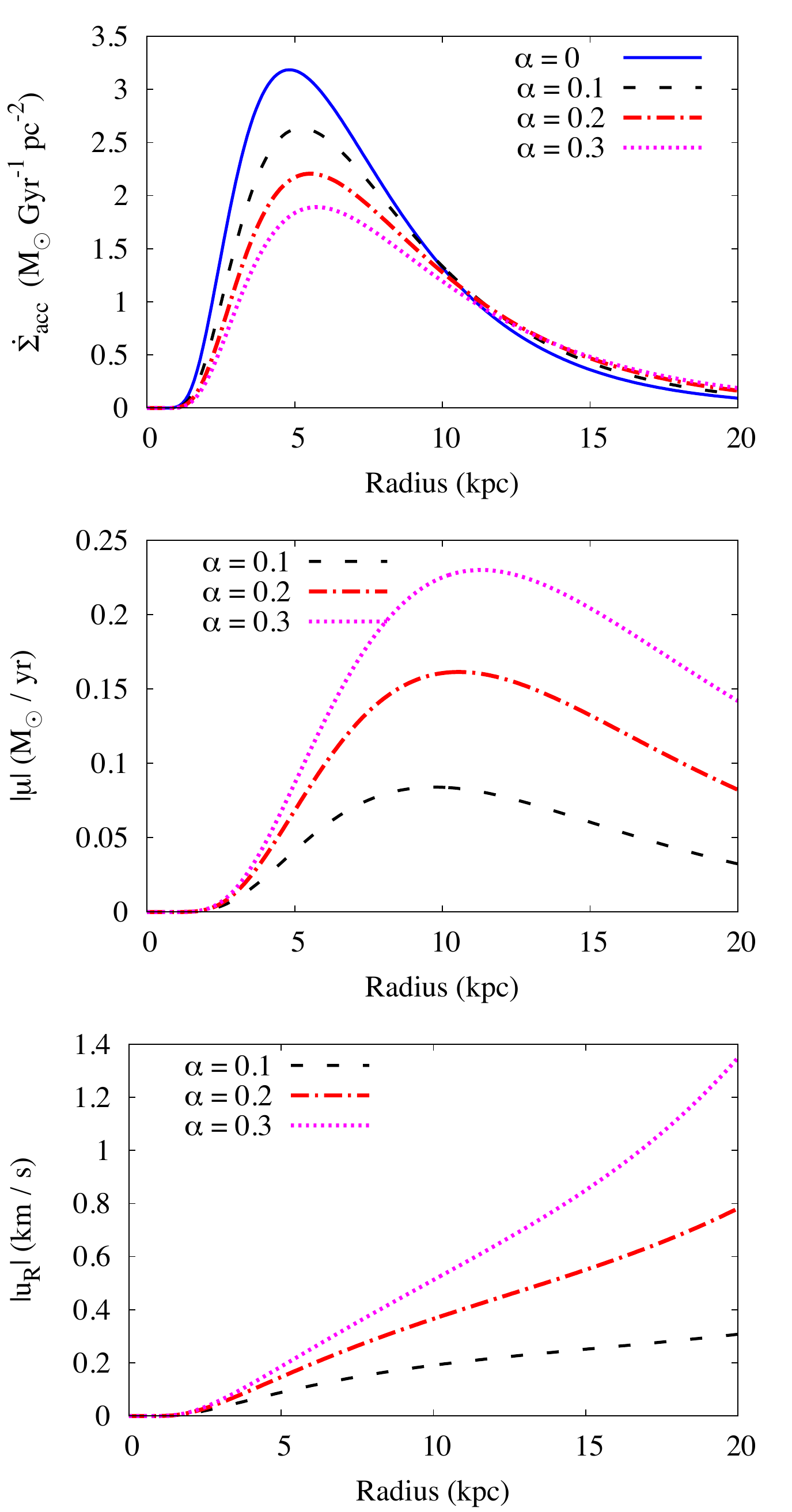}
\caption{Similar to Fig. \ref{fig::FluxPlot}, but for an inside-out model of the Galaxy (Sec. \ref{sec::InsideOutParameters}). Note that a linear scale is adopted here for all the plots. The accretion profile has a marked peak at a finite radius, like the profile of star formation (Fig. \ref{fig::SFRD}). With increasing $\alpha$, the peak of accretion moves outwards and progressively strong radial flows develop at large radii, bringing the gas, accreted in the outskirts with low angular momentum, towards the inner star formation peak.}\label{fig::InsideOutFlux}
\end{figure}
In Fig. \ref{fig::InsideOutFlux}, the mass flux decomposition, computed as described in Sec. \ref{sec::MassFluxDec}, is reported as a function of the angular momentum mismatch parameters $\alpha$, in a similar fashion to Fig. \ref{fig::FluxPlot}, but now for the inside-out growing model derived above.

The upper panel shows the predicted present-day accretion profiles for different values of $\alpha$. The first thing to be noticed is that accretion in this case always peaks at a finite galactocentric radius (at variance with the non inside-out case, where it always had its maximum at the origin, cfr. Fig. \ref{fig::FluxPlot}, upper panel). This is a distinctive feature of an inside-out growing model, which is also clearly reflected in the observed SFRD profile (Fig. \ref{fig::SFRD}). Furthermore, the radius where maximum accretion occurs is an increasing function of $\alpha$. This is because, when some local angular momentum deficit is taken into account, accretion preferentially occurs at large radii, from which fresh gas is then brought inwards by radial flows.

The inwards radial mass flux $(-\mu)$, consistently predicted by the mass-flux decomposition, is shown in the middle panel of Fig. \ref{fig::InsideOutFlux}. According to the general property pointed out in Sec. \ref{sec::FluxDecompositionApplication}, the position of the peak of $(-\mu)$ coincides with the radius outside which the predicted accretion is enhanced with respect to the effective one. The peak radius is larger now with respect to the non inside-out growing case: radial flows in this model bring inwards material that has been accreted, with relatively low angular momentum, at very large radii.

Finally, in the lower panel of Fig. \ref{fig::InsideOutFlux} the predicted present-day patterns are reported for the inward radial velocity of gas, as a function of radius, for different values of $\alpha$. Again, these show a non-trivial trend, which is also strongly varying with $\alpha$ and it is different from the one predicted in the non inside-out model (Fig. \ref{fig::FluxPlot}), especially in the inner and outer regions. This indicates that each particular structural evolution model requires its own mass flux decomposition, for radial flows to be compatible with angular momentum conservation; such a decomposition can always be easily achieved by means of the explicit formulae given in Sec. \ref{sec::MassFluxDec}.

\subsection{Abundance gradients in inside-out models with radial flows}
We finally apply the methods described in Sec. \ref{sec::ChemicalEvolutionRadialFlows} to compute the abundance gradients in an inside-out model in the presence of accretion-induced radial flows. We adopt the domain-extension strategy and we assume $\alpha$ to be constant with both space and time. 

A technical digression is due here for completeness. Since our inside-out model has a formally vanishing gas density at the disc birth, then equation \eqref{defmu} formally implies infinite radial velocities at $t = 0$, which fact is obviously unphysical. \cite{PT96} studied this problem with hydrodynamical simulations and concluded that, when very strong accretion occurs on to a disc with very low surface density, centrifugal equilibrium is rapidly, but not immediately, achieved, with the consequence that the physically motivated radial mass flux is smaller than the one predicted by \eqref{MVeq}. In our models, these conditions occur only at very early times and at very large radii and are thus expected to have a very limited impact (see also \citealt{ChamchamTayler94}). We performed several experiments, adopting different strategies to ensure a physically meaningful velocity field, and we verified that the results that we provide in the following are insensitive to the details of the adopted scheme.

\begin{figure}
\centering
\includegraphics[width=9cm]{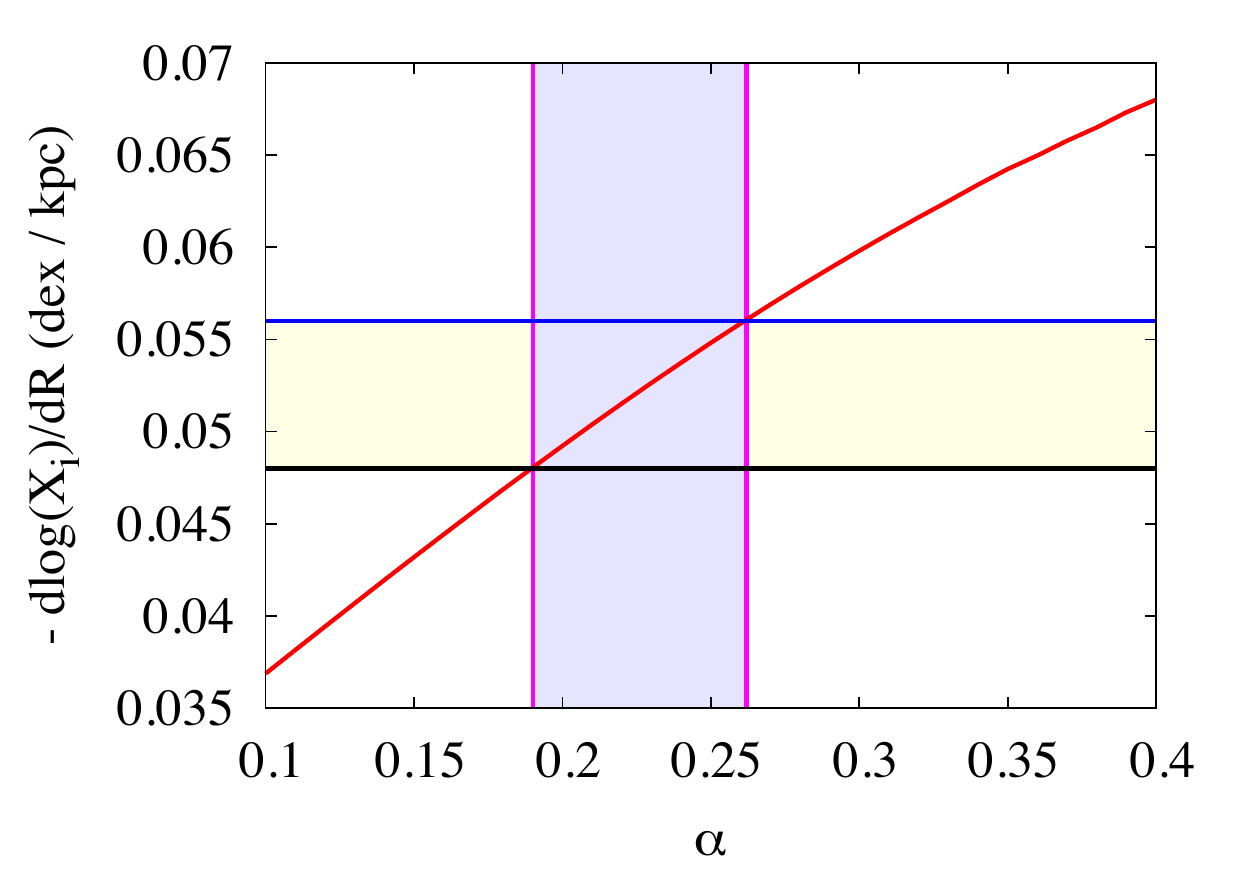}
\caption{Similar to Fig. \ref{fig::GradPlot}, but for an inside-out model of the Galaxy (Sec. \ref{sec::InsideOutParameters}). Slightly lower values of $\alpha$ are needed in this case to explain the observed abundance gradients, corresponding to accreting material rotating with as much as 81 per cent of the rotational velocity of the disc.}\label{fig::GradPlotInsideOut}
\end{figure}

As in Sec. \ref{sec::Calibration}, we compare the model predictions with the observed present-day $\alpha$-element abundance gradient in Cepheid stars. Fig. \ref{fig::GradPlotInsideOut}, which is the analogue of Fig. \ref{fig::GradPlot}, reports this comparison for the inside-out case. With respect to the previous model, the model with inside-out growth predicts a steeper gradient for any given value of $\alpha$. This is because inside-out growth itself has the well known effect of steepening abundance gradients. As a consequence, the range for the parameter $\alpha$ inferred from observations is shifted towards lower, but still non-vanishing, values for the angular momentum mismatch parameter. The derived interval is $0.19 \leq \alpha \leq 0.26$, corresponding, in the approximation of a flat-rotation curve for the disc, to cold gas accreting on the disc of the Galaxy with $74 - 81$ per cent of the rotational velocity of the disc. Also this finding is in very good agreement with \cite{BS12}, despite their very different approach to inside-out growth, tentatively pointing towards the robustness of the result.

\section{Summary and conclusions}\label{sec::Summary}
The gas accreting on the discs of spiral galaxies is likely to rotate, on average, at a lower speed with respect to the disc itself. This is especially true if the accretion comes from a hot reservoir, which must rotate more slowly than the disc, because of its partial pressure support against gravity. A similar effect is also likely for cold-mode accretion, on geometrical grounds. The local angular momentum mismatch in accretion drives inward radial flows within the disc, due to angular momentum conservation, with at least two important consequences on structural and chemical evolution models. First, the accretion profile, as indirectly inferred from the observed structure of discs, is severely altered, in the sense of an enhancement of the required accretion in the outer regions and a corresponding depletion in the inner ones. Secondly, the spatial distribution of heavy elements is profoundly affected, in the sense of the creation or strong steepening of abundance gradients. Abundances are observable and can be used to shed light on accretion and radial flows, which are still not. In this work, we addressed the computation of accretion profiles, radial flows and abundance gradients, as a function of the angular momentum of the accreting material, from an analytic point of view.

The advantages of an analytic treatment come at the cost of simplifying assumptions. In particular, we assumed metallicity-independent yields and the instantaneous recycling approximation; we also did not address stellar radial migration and mechanisms of radial mixing other than those due to angular momentum conservation. These simplifications limit the predictive power of our chemical analysis to the abundance profile of $\alpha$-elements in the ISM. However, this observable is so remarkably sensitive to the angular momentum of the infalling material that it is very useful to extract the associated information by means of a simple and easily controllable machinery.

Our work can be summarized as follows.

\begin{enumerate}
\item We considered the simplest possible model for the evolution of a galaxy disc -- namely an exponential disc, with constant scalelength, obeying the Kennicutt-Schmidt law -- and we have shown that, as long as radial gas flows are neglected, it is incompatible with the observed abundance gradients of spiral galaxies.
\item We provided the general exact solution to the problem of decomposing the gas flows, needed to sustain galaxy evolution, into direct accretion from the IGM and radial flows within the disc (equations \eqref{GeneralSolution} to \eqref{uRtrivial}), which can be applied to any axisymmetric, non-viscous model for the structural evolution of a disc and to any angular momentum distribution of the accreting material; we emphasize that with this approach both the accretion profile and the radial velocity pattern are model predictions, rather than model assumptions.
\item We proposed a novel method, based on characteristic lines, to solve the equation of metallicity evolution in the presence of radial flows, which makes the computation of gas-phase $\alpha$-element abundance profiles very easy and allows the influence of boundary conditions on the steepness of outer gradients to be traced straightforwardly.
\item We suggested a strategy to disentangle the contributions of inside-out growth and radial flows in determining the steepness of abundance gradients: inside-out growth parameters can be fixed to those required to reproduce non-chemical observables and in particular the shape of the SFRD profile, which is particularly sensitive to the structural evolution of the disc; the contribution of inside-out growth to abundance gradients is thus fixed and residual discrepancies with observations, if present, can then be used to constrain the needed amount of radial flows.
\end{enumerate}

Given their simplicity and generality, these methods can be readily applied to any future model of the evolution of our Galaxy or of external spiral galaxies, provided that radial flows are dominated by angular momentum conservation.

Simple illustrative models, calibrated on the Milky Way, require, to match observations, a rotational velocity for the accreting material comprised between $70$ and $80$ per cent of the rotational velocity of the disc, in excellent agreement with previous estimates. Once this effect is taken into account, a picture emerges in which a significant part of the accretion of cold gas occurs in the outskirts of the disc and it is then brought inwards by radial flows, towards the inner regions where it is required to sustain star formation.

\section*{Acknowledgements}
The authors would like to thank R. Sch\"onrich and D. Romano for useful discussion. They also acknowledge financial support from PRIN MIUR 2010-2011, project `The Chemical and Dynamical Evolution of the Milky Way and Local Group Galaxies', prot. 2010LY5N2T.

\bibliographystyle{mn2e}
\bibliography{mybib}{}

\appendix
\section{Angular momentum and radial flows in fountain-driven accretion}
We briefly sketch here the basic ingredients for the study of accretion-induced radial flows in the context of the theory of fountain-driven accretion (\citealt{Marinacci+10}; \citealt{Marinacci+11}; \citealt{Marasco+12}).

\begin{figure}
\centering
\includegraphics[width=8cm]{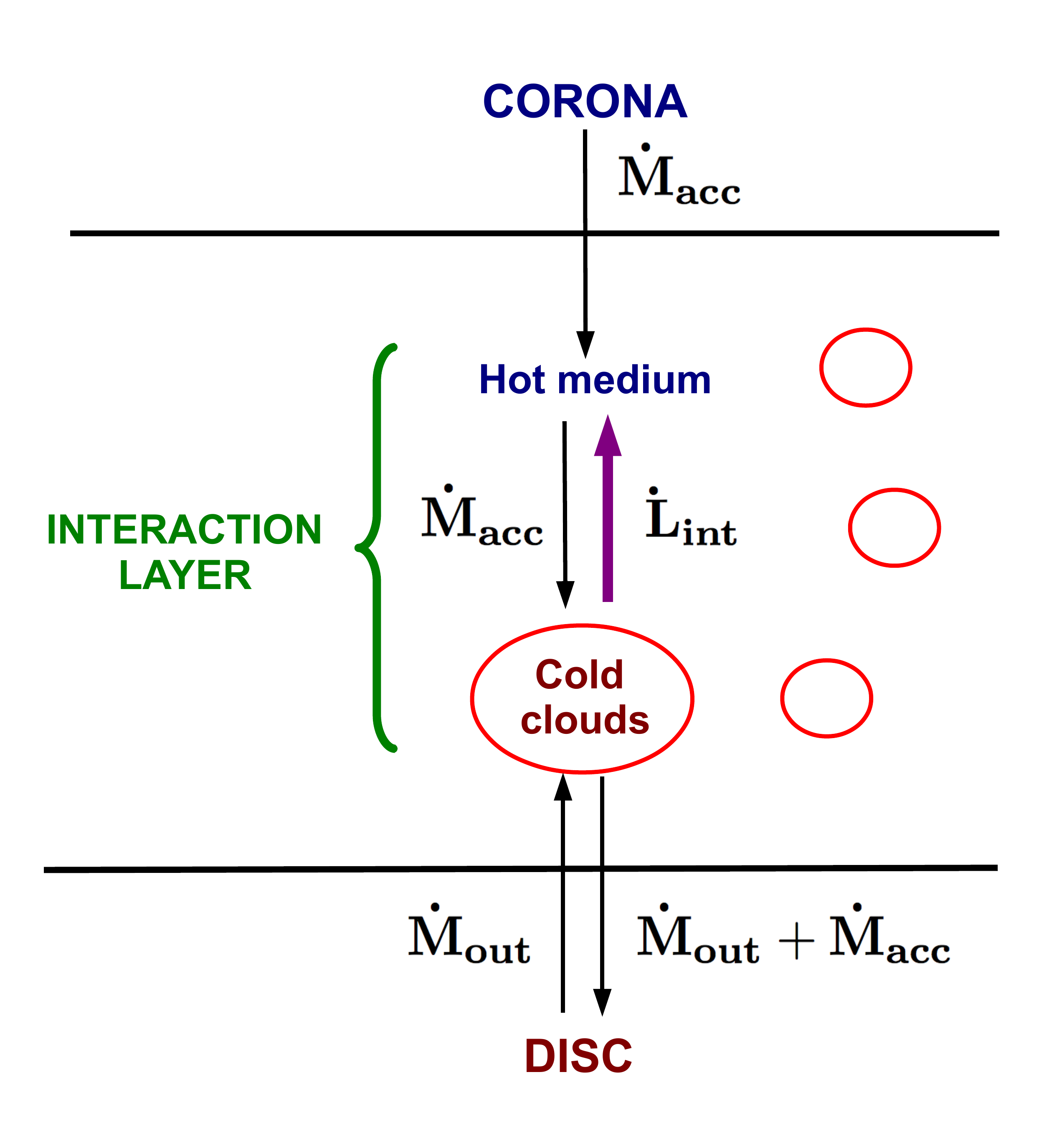}
\caption{Schematic diagram of the accretion of coronal material on to the disc, according to the theory of Marinacci et al. (2011). After an initial transfer of angular momentum, the interaction with cold fountain clouds triggers the condensation of a part of the hot medium.}\label{fig::Scheme}
\end{figure}

Fig. \ref{fig::Scheme} is a diagram of the involved phases and processes. Three regions are considered: the disc, the (upper) corona and an intermediate layer, where the interaction between the two occurs, the thickness of which is set by the vertical extent of the galactic fountain. In the intermediate layer, we distinguish two spatially mixed phases: a cold clumpy phase, constituted of fountain clouds together with coronal material condensing on them (this phase is observed in HI as extraplanar gas), and a hot diffuse phase, which is the lower part of the corona, in direct contact and interaction with the cold clouds. Thin black arrows indicate mass exchanges between different phases; to each of them a (not shown) advective angular momentum exchange is also associated, equal to the mass flux multiplied by the specific angular momentum of the upstream region: for instance, the net angular momentum gain of the disc is $\dot{L}_\textrm{disc} = (\dot{M}_\textrm {out}+\dot{M}_\textrm{acc})l_\textrm{cold} - \dot{M}_\textrm{out}l_\textrm{disc}$. In addition, there is a non advective angular momentum exchange, marked with a thick magenta arrow, between the cold and the hot phase in the intermediate layer: this is predicted by the  \cite{Marinacci+11} theory and its outcome is to set an equilibrium angular momentum difference between the cold and the hot phase:
\begin{equation}\label{MarinacciEq}
 l_\textrm{cold}-l_\textrm{hot} = (\Delta l)_\textrm{eq}
 \end{equation}
which allows the condensation of coronal gas on fountain clouds to become effective. Here $(\Delta l)_\textrm{eq} = R (\Delta V)_\textrm{eq}$, where $R$ is the galactocentric radius of the considered region and $(\Delta V)_\textrm{eq} \simeq 75 \; \textrm{km}/\textrm{s}$ is the equilibrium velocity lag predicted by the theory. The mass flux from the upper corona to the lower corona is the one required to balance mass loss suffered by the hot phase as a consequence of fountain-driven accretion (in practice, it is driven by the imbalance between pressure and gravity due to the disappearance of part of the underlying hot material). According to \cite{FB08}, the angular momentum exchange between the cold and the hot phase is very efficient and it brings the system towards an approximate equilibrium in few dynamical times. During this time, we may assume that the fountain ejection rate $\dot{M}_\textrm{out}$ does not change very much, nor the mass reservoir of the upper corona will be significantly eroded and therefore an approximately stationary state is a reasonable assumption on such time-scales. Angular momentum balance for the hot and cold phases therefore give:
\begin{equation}
\left\{ \begin{array}{l}
\dot{M}_\textrm{acc} l_\textrm{corona} + \dot{L}_\textrm{int} = \dot{M}_\textrm{acc}l_\textrm{hot}\\
\dot{M}_\textrm{out} l_\textrm{disc} + \dot{M}_\textrm{acc} l_\textrm{hot} = \dot{L}_\textrm{int} + (\dot{M}_\textrm{out} + \dot{M}_\textrm{acc}) l_\textrm{cold}
\end{array}\right.
\end{equation}
which, together with equation \eqref{MarinacciEq}, can be solved for $l_\textrm{cold}, l_\textrm{hot}, \dot{L}_\textrm{int}$. In particular, we find:
\begin{equation}\label{lcold}
l_\textrm{cold} = \frac{\dot{M}_\textrm{out} l_\textrm{disc} + \dot{M}_\textrm{acc} l _\textrm{corona}}{\dot{M}_\textrm{out} + \dot{M}_\textrm{acc}}
\end{equation}

Note that the angular momentum of the cold accreting gas does not depend on the equilibrium lag $(\Delta l)_\textrm{eq}$, which, however, sets the rotation velocity of the lower corona and the magnitude of non-advective angular momentum exchanges within the interacting layer.

In the approximation of a flat rotation curve \eqref{alphaMestel}, the angular momentum mismatch parameter reads:
\begin{equation}\label{alphacold}
\alpha_\textrm{cold} \vcentcolon= 1 - \frac{l_\textrm{cold}}{l_\textrm{disc}} = \frac{\eta}{1 + \eta} \alpha_\textrm{corona}
\end{equation}
with obvious meaning of $\alpha_\textrm{corona}$, while
\begin{equation}\label{etadef}
\eta \vcentcolon= \dot{M}_\textrm{acc}/\dot{M}_\textrm{out}
\end{equation}
depends on the condensation efficiency and on the orbital time of the fountain clouds. Note that $\eta$ can vary with position within the disc. More in general, all the discussion above is radial dependent and all masses and angular momenta should be accordingly translated into the corresponding surface densities.

According to equation \eqref{MVeq}, the radial mass flux is therefore given by:
\begin{equation}\label{MVtot}
\mu = - 2 \pi  R^2 \alpha_\textrm{cold} (\dot{\Sigma}_\textrm{out} + \dot{\Sigma}_\textrm{acc})
\end{equation}
which, considering equations \eqref{alphacold} and \eqref{etadef}, can also be written:
\begin{equation}\label{finaleq}
\mu = - 2 \pi  R^2 \alpha_\textrm{corona} \dot{\Sigma}_\textrm{acc}
\end{equation}

Therefore, despite the fact that cold material accretes on to the disc with a specific angular momentum that can be very different from the one of the upper corona (equation \eqref{alphacold}), the induced radial flows within the disc are nonetheless the same that one would have if the coronal material was accreting on to the disc directly (equation \eqref{finaleq}), with the galactic fountain basically acting as a catalyst for accretion. 

Of course, this conclusion is just a first approximation, the validity of which is limited by our numerous simplifying assumptions. Among them, we underline that our basic diagram (Fig. \ref{fig::Scheme}) can be read locally (as we did) only provided that the orbital radial excursion of each fountain is small compared with its starting radius. While this is nearly correct for moderate ejection velocities (\citealt{FB06}; \citealt{SRM08}), it will not be true in detail (\citealt{Marasco+12}) and it will fail for a very powerful fountain, so that some level of distortion to equation \eqref{finaleq} is indeed expected in general, the magnitude of which requires a dedicated study. The key prediction remains, however, that the main regulator for the radial flows in the disc has to be looked for in the angular momentum distribution of the upper corona, which is the ultimate source of mass and angular momentum in the model, but of which we still have a rather scant knowledge.

\end{document}